\documentclass[structabstract]{aa}
\usepackage{graphicx}
\usepackage{lscape}
\def\approxgt{\mathrel{\hbox{\rlap{\lower.55ex \hbox {$\sim$}}
        \kern-.3em \raise.4ex \hbox{$>$}}}}
\def\approxlt{\mathrel{\hbox{\rlap{\lower.55ex \hbox {$\sim$}}
        \kern-.3em \raise.4ex \hbox{$<$}}}}
\begin{document}
   \title{The X-ray view of Giga-Hertz
          peaked spectrum radio galaxies}

   \author{Olof Tengstrand
          \inst{1,2},
          M.Guainazzi
          \inst{1},
	  A.Siemiginowska
          \inst{3},
          N.Fonseca Bonilla
          \inst{1},
	  A.Labiano
	  \inst{4},
	  D.M.Worrall
          \inst{5},
          P.Grandi,
          \inst{6}
	  E.Piconcelli
          \inst{7}
          }

   \offprints{M.Guainazzi}

   \institute{$^1$European Space Astronomy Centre of ESA P.O.Box 78,
              Villanueva de la Ca\~nada, E-28691 Madrid, Spain \\
	      \email{Matteo.Guainazzi@sciops.esa.int} \\
              $^2$Institute of Technology, University of Link\"oping,
	      SE-581 83, Link\"oping, Sweden \\
	      $^3$Harvard-Smithsonian Centre for Astrophysics, 60 Garden St.,
		Cambridge, MA 02138, USA \\
	      $^4$Departamento de Astrof\'isica Molecular e Infrarroja,
		Instituto de Estructura de la Materia (CSIC), Madrid, Spain \\
	      $^5$H. H. Wills Physics Laboratory, University of Bristol, 
		Tyndall Avenue, Bristol BS8 1TL, United Kingdom \\
 	      $^6$Istituto di Astrofisica Spaziale e Fisica Cosmica-Bologna,
		INAF, via Gobetti 101, I-40129 Bologna\\
	      $^7$Osservatorio Astronomico di Roma (INAF), via Frascati 33,
		00040 Monteporzio Catone, Roma, Italy \\
              }

   \date{Received ; accepted }

   \abstract{This paper presents the X-ray properties of a flux- and
volume-limited complete sample of 16 Giga-Hertz Peaked Spectrum (GPS) galaxies.}
	    {This study addresses three basic questions in our understanding
of the nature and evolution of GPS sources: a) What is the physical origin of
the X-ray emission in GPS galaxies? b) What physical system is associated
with the X-ray obscuration? c) What is the ``endpoint'' of the evolution of
compact radio sources?}
	    {We discuss the results of the X-ray spectral 
analysis, and compare the X-ray properties of the sample sources with radio
observables.}
	    {We obtain a 100\% (94\%) detection fraction in the 0.5--2~keV
(0.5--10~keV) energy band. GPS galaxy X-ray spectra are typically highly
obscured ($\langle N_H^{{\rm GPS}} \rangle = 3 \times 10^{22}$~cm$^{-2}$;
$\sigma_{N_H} \simeq 0.5$~dex).  The X-ray column density is higher than the
HI column density measured in the radio by a factor of 10 to 100. GPS galaxies
lie well on the extrapolation to high radio powers of the correlation between
radio and X-ray luminosity known in low-luminosity FR~I radio galaxies. On the
other hand, GPS galaxies exhibit a comparable X-ray luminosity to FR~II radio
galaxies, notwithstanding their much higher radio luminosity.}
	    {The X-ray to radio luminosity ratio distribution in our sample is
consistent with the bulk of the high-energy emission being produced by the
accretion disk, as well as with dynamical models of GPS evolution where X-rays
are produced by Compton upscattering of ambient photons. Further support for
the former scenario comes from the location of GPS galaxies in the X-ray to
O[{\sc iii}] luminosity ratio versus $N_H$ plane. We propose that GPS galaxies
are young radio sources, which would reach their full maturity as classical
FR~II radio galaxies. However, column densities $\approxgt$10$^{22}$~cm$^{-2}$
could lead to a significant underestimate of dynamical age determinations
based on the hotspot recession velocity measurements.}

   \keywords{Galaxies: jets --
	     Galaxies: active --
 	     X-ray: galaxies
            }

\authorrunning{Tengstrand et al.}

\titlerunning{X-ray view of GPS radio galaxies}

\maketitle
%
%________________________________________________________________

\section{The nature of GPS radio galaxies}

This paper presents an X-ray study
of a complete radio-selected sample of giga-hertz
peaked spectrum (GPS) galaxies.
GPS sources are characterised
by a simple convex radio spectrum peaking
near 1 GHz
(Stanghellini 2006, Lister 2003, O'Dea 1998). They represent
about 10\% of the 5-GHz selected sources. About half
of the known GPS sources are
morphologically classified as galaxies, the remaining
as quasars.
They
often exhibit symmetric, very compact (10--100~pc)
structures,
reminiscent of those present in extended radio galaxies
on much larger scales.

Little is known about their high-energy emission.
GPS galaxies are rather elusive in X-rays (O'Dea et al. 1996).
X-ray spectroscopic studies prior to modern X-ray
observatories were inconclusive on
whether this low detection rate is due to intrinsic
weakness or to obscuration of the active nucleus (\cite{elvis94a}).
Deep {\it Chandra} and XMM-Newton observations of GPS galaxies
are scanty. One of the few exceptions is a deep
XMM-Newton pointing of 3C301.1 (\cite{odea06}); it
revealed a hard X-ray emission component, which could
be associated with hot gas shocked by the expansion of the
radio source or to synchrotron self-Compton emission.
Analysis of small samples of GPS galaxies observed
with XMM-Newton were presented by Vink et al. (2005) and
Guainazzi et al. (2006). Our paper represents an
extension of their results.

Understanding the origin of high-energy emission
in these objects may have important implications
for the birth and evolution of the ``radio power'' in the
Universe. 
GPS sources were originally suggested to represent
radio galaxies in the early stage of their life
(typical ages $< 10^4$~years;
Fanti et al. 1995; Murgia 2003). This possibility was recently
supported by the detection of
hotspot proper motions
(Poladitis \& Conway 2003; Gugliucci et al. 2005).
Alternatively, as originally suggested 
by Gopal-Krishna \& Wiita (1991), GPS
sources could remain
compact during their whole radiative lifetime,
because interaction with dense
circumnuclear matter impedes
their full growth.

In order to address the above issues, and provide
the best possible estimate of the gas density
in the GPS galaxy nuclear environment,
we have undertaken an XMM-Newton observation program
of a radio-selected complete sample of GPS galaxies.
The three main issues
which originally motivated our study, that
will be discussed throughout the paper, are:

\begin{itemize}

\item What is the physical origin of the X-ray emission
in GPS galaxies?

\item Which physical system is associated with the X-ray obscuration?

\item What is the ``endpoint'' of the evolution of
compact radio sources?

\end{itemize}

This paper is structured as follows: in Sect.~2 we present
the sample. The data reduction procedure and the
results of the spectral analysis on
the unpublished sources are presented
in Sect.~3 and 4, respectively. This paper does
not discuss the X-ray spectral analysis of
sources published elsewhere (\cite{guainazzi06,vink05}).
In Sect.~5 we summarise the X-ray properties of the whole
GPS sample, and compare these with a control sample
of large-scale radio galaxies. Finally, we discuss 
our results in Sect.~6, and summarise our findings
and their possible implications for the evolution
of GPS sources in Sect.~7.
We adopt throughout the paper a cosmology with
$H_0 = 70$~km~s$^{-1}$~Mpc$^{-1}$, $\Omega_{\Lambda}$=0.73,
and $\Omega_m$=0.27 (\cite{spergel07}).

\section{The sample}

The sources discussed in this paper
constitute a flux- and volume-limited sub-sample
extracted from the complete radio-selected sample of GPS galaxies of
Stanghellini et al. (1998). We refer to this
sub-sample as ``our GPS sample'' hereafter.
We selected all sources with redshift
$z$$<$1, and flux density at 5~GHz $\ge$1~Jy.
The whole sample
(16 sources) has been observed with XMM-Newton across different
observing cycles. The only exceptions are
PKS~0941$-$08 and PKS1345+125, for which archival {\it
Chandra} and ASCA data are available, respectively.
Preliminary results, based on a small
sub-sample of 5~objects, were presented in Guainazzi et al.
(2006)\footnote{Guainazzi et al. (2006) present also
data of COINSJ0029+3456; this source was later discovered to
host a blazar, and will not be considered in the sample
discussed in this paper.}.
Some of the sample sources were included in Vink et al. (2005).
The whole sample is listed in Table~\ref{tab1}.
%--------------------- Table 1
\begin{table*}
\begin{small}
\caption{\small The GPS sample discussed in this paper.}
\label{tab1}
\end{small}
\begin{footnotesize}
\begin{center}
\begin{tabular}{lccccc}  \hline \hline
NED name    & J2000 name & z$^a$ & Observation ID$^b$ & Observation date  &Reference$^c$ \\ \hline
4C+00.02        & 002225+001456 & 0.305 & 0407030101 & 2006-Jun-26  &(1)\\
COINSJ0111+3906 & 011137+390628 & 0.668 & 0202520101 & 2004-Jan-09  &(2)\\
PKS0428+20      & 043103+203734 & 0.219 & 0407030201 & 2006-Sep-02  &(1)\\
PKS0500+019     & 050321+020305 & 0.585 & 0205180601 & 2004-Aug-18  &(3)\\
B30710+439      & 071338+434917 & 0.518 & 0202520201 & 2004-Mar-23  &(2)\\
PKS0941$-$080   & 094336$-$081931 & 0.228 & Chandra    &              &(3)\\
B1031+567       & 103507+562847 & 0.450 & 0202520301 & 2004-Oct-21  &(2)\\
4C+14.41        & 112027+142054 & 0.362 & 0502510201 & 2007-Jun-13  &(1)\\
4C+32.44        & 132616+315409 & 0.370 & 0502510301 & 2007-Dec-05  &(1)\\
PKS1345+125     & 134733+121724 & 0.122 & ASCA       &              &(4)\\
4C+62.22        & 140028+621038 & 0.431 & 0202520401 & 2004-Apr-14  &(2)\\
OQ+208          & 140700+133438 & 0.077 & 0140960101 & 2003-Jan-31  &(5)\\
PKS1607+26      & 160913+264129 & 0.473 & 0502510401 & 2008-Jan-17  &(1)\\
                &               &       & 0502510801 & 2008-Jan-19  &   \\
PKS2008$-$068   & 201114$-$064403 & 0.547 & 0502510501 & 2007-Oct-12  &(1)\\
PKS2127+04      & 213032+050217 & 0.990 & 0502510601 & 2007-May-17  &(1,6,7)\\
COINSJ2355+4950 & 235509+494008 & 0.238 & 0202520501 & 2003-Dec-26  &(2)\\
\hline \hline
\end{tabular}
\end{center}
$^a$ Data taken from NED \\
$^b$ XMM-Newton observation ID \\
$^c$ (1)~this paper; (2)~\cite{vink05}; (3)~\cite{guainazzi06};
(4)~\cite{odea00}; (5)~\cite{guainazzi04} ; (6)~\cite{siemiginowska05};
(7)~\cite{siemiginowska08}
\end{footnotesize}
\end{table*}
%------------------------- Table 1
A more general discussion of the X-ray and multiwavelength properties of
our whole GPS sample is deferred to Sect.~5.
A summary of the X-ray and radio properties of the whole sample
is in Appendix~B.

\section{Observations and data reduction}

In this paper [as also originally done by Guainazzi et al. (2006)
and Vink et al.
(2005)] we will consider only X-ray data taken with the XMM-Newton EPIC
cameras (pn, \cite{struder01}; MOS, \cite{turner01}), because the
sources were too faint to yield a measurable signal in the high-resolution
RGS cameras.
Observational information for the sources for which
new measurements are presented here is listed in Table~\ref{tab3}.
%--------------------- Table 2
\begin{table}
\begin{small}
\caption{\small Properties of the X-ray observations discussed in this paper.
Exposure times (T$_{exp}$) and Count Rates ($CR$) refer to the pn
in the 0.5--10~keV band (unless otherwise specified).
$\tau$ is the count rate threshold applied in the determination
of the Good Time Interval for scientific product extraction (details in
text). $\rho$ is the radius of the scientific products extraction
region. $N_{H,Gal}$ is the column density due to gas in the Milky Way
along the line-of-sight to the GPS galaxy (\cite{kalberla05})
in units of $10^{20}$~cm$^{-2}$.}
\label{tab3}
\end{small}
\begin{footnotesize}
\begin{center}
\begin{tabular}{lccccc} \hline \hline
Source             & N$_{H,Gal}$ & T$_{exp}$ & $\tau$ & $\rho$ & $CR$ \\
& & (ks) & (s$^{-1}$) & (\arcsec) & (10$^{-3}$~s$^{-1}$) \\ \hline
4C+00.02        & 2.7 & 22.2 & 2.1        & ...            & $^a$ \\
PKS0428+20      & 19.6 & 15.4 & 0.75       & 23             & $7.1 \pm 0.8$ \\
4C+14.41        & 2.0 & 16.3 & 0.5        & 15             & $4.8 \pm 0.8$ \\
4C+32.44        & 1.2 & 20.5 & 0.5        & 16             & $25.6 \pm 1.2$ \\
PKS1607+26      & 3.8 & 18.5 & 18         & 20             & $39  \pm 7 $ \\
PKS2008$-$068     & 5.0 & 23.2 & 0.75       & 15             & $2.0 \pm 0.4$ \\
PKS2127+04      & 5.0 & 10.8 & 0.5        & 25             & $21.2  \pm 1.9 $ \\
\hline
\end{tabular}
\end{center}
\end{footnotesize}
\noindent
$^a$detected only in the soft X-ray band (0.5--2~keV), with
a count rate: $(2.3 \pm 0.7) \times 10^{-3}$~s$^{-1}$.
\end{table}
%------------------------- Table 2

XMM-Newton data were reduced with SASv7.1 (\cite{gabriel03})
according to standard procedures as in, {\it e.g.},
Guainazzi et al. (2006). The most updated calibration files available at 
the date of the analysis (February 2008) were used. 
Source scientific products were accumulated from circular regions 
surrounding the position of the optical nucleus of each source
(extracted from the NED catalogue)\footnote{{\tt
http://nedwww.ipac.caltech.edu/}}. The size of the
source extraction regions are shown in Table~\ref{tab3}.
Following Guainazzi (2008), background scientific products
were extracted from source free circular regions
close to the source and on the same CCD as the source
for the MOS cameras; and from source-free regions
centred at the same
row in detector coordinates as the source in nearby CCDs
for the pn. As many of
the sources were X-ray faint,
particular care has been applied in the choice of
flaring particle background rejection
thresholds optimising the signal-to-noise ratio of the final scientific
products. Using a single-event, $E>$10~keV full field-of-view light
curve as a monitoring tool of the instantaneous intensity of the
background, ten different logarithmically spaced thresholds
between 0.1~$s^{-1}$ and about two times the highest
light curve count rates were tried.
For each threshold the radius of the source extraction region
was also varied to obtain the highest number of net counts for a given
signal-to-noise ratio.  
Spectra were binned in such a way as to avoid oversampling
of the intrinsic instrumental energy resolution by
a factor larger than 3, and to have at least 25 background-subtracted
counts in each spectral bin. These conditions ensure
the applicability of the $\chi^2$ statistic as a goodness-of-fit test.

In this paper, errors on the spectral parameters
and on any derived quantities are at the 90\% confidence
level for one interesting parameter; errors on the count rates and
derived quantities are at 1$\sigma$ level. Whenever statistical
moments or correlations on distributions including upper limits are
calculated, an extension of the regression method on censored data
originally described by Schmitt (1985) and Isobe et al.
(1986) has been used.
More details on this method were presented by Guainazzi et al. (2006).

\section{Results}

In this Section we present spectral-analysis for
the 7 unpublished sources in our GPS sample.

No significant variability in either integrated X-ray flux or spectral
shape was detected in any source presented in this paper
on timescales $\approxlt$10$^4$~s.
We therefore focus on
the properties
of their time-averaged spectra.

\begin{figure*}
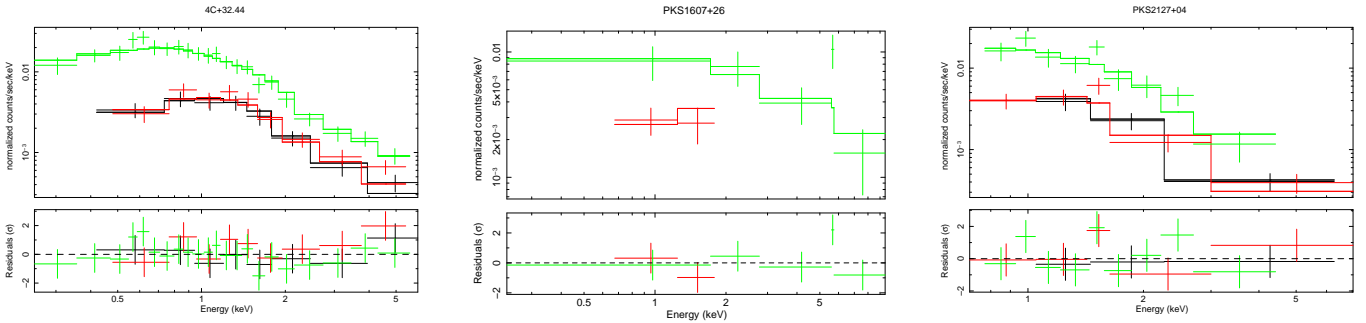

\begin{center}
\hbox{
\includegraphics[height=55mm,angle=270]{./4C3244.cps}
\hspace{0.5cm}
\includegraphics[height=55mm,angle=270]{./PKS1607.cps}
\hspace{0.5cm}
\includegraphics[height=55mm,angle=270]{./PKS2127.cps}
}
\end{center}
\caption{Spectra of the three EPIC instruments
({\it upper panels}), and
residuals in units of standard deviation
({\it lower panels}) when the baseline model is applied.
In addition to the binning applied to the spectra to ensure
the applicability of the $\chi^2$ goodness-of-fit test,
the spectra were further rebinned in this Figure such that
each spectral channel corresponds to a signal-to-noise ratio
$>$3 for plotting purposes only. {\it Green}: pn; {\it black}: MOS~1;
{\it red}: MOS~2.
}
\label{fig:Spectra}
\end{figure*}

For 3 of these
sources the number of degrees of freedom in the binned spectra
was larger than 4: 4C+32.44, PKS1607+26, PKS~2127+04.
In these cases a standard spectral analysis was possible. The 
spectra were fitted in the 0.2--10~keV energy range with
{\sc Xspec} (version 11; \cite{arnaud96}).
A model consisting of three
components was used:     
\begin{eqnarray}
e^{-N_{H,Gal} \sigma(E)} \times e^{{-N_H} \sigma [E(1+z)]} \times KE^{-\Gamma}
\end{eqnarray}
\noindent
where the photo-electric absorption components use
Wisconsin cross-sections (\cite{morrison83}), and $K$
is the unabsorbed spectral normalisation at 1~keV.
We'll refer to this model as our ``baseline'' model
hereafter.
The column density $N_{H,Gal}$ was kept fixed to the
contribution by intervening gas in the Galaxy as measured in the
Leiden/Argentine/Bonn (LAB) Survey of Galactic HI
(\cite{kalberla05})\footnote{http://heasarc.gsfc.nasa.gov/cgi-bin/Tools/w3nh/w3nh.pl},
whereas $N_H$ is the column density of local gas at the
galaxy redshift.
The results of the spectral fitting are
shown in Fig.~\ref{fig:Spectra} and summarised in Table~\ref{tab2}.
%-------------- Tab.4
\begin{table*}
\begin{small}
\caption{\small Best-fit parameters and results for the sources
where spectral analysis was possible.}
\label{tab2}
\end{small}
\begin{footnotesize}
\begin{center}
\begin{tabular}{lccccccc} \hline \hline
Source         & $N_H$                   &$\Gamma$                 & F$_{2-10}$        & $\log (L_{2-10}^a)$ & Fe K$_{\alpha}$~EW  &$\chi ^2/\nu$ \\ 
            & (10$^{21}$~cm$^2$)       &                         &(10$^{-13}$~ergs~cm$^{-2}$~s$^{-1}$) & & (keV) &  \\ \hline
4C+32.44    & $1.2 \pm^{0.6}_{0.5}$ & $1.74 \pm 0.2$  & $0.78 \pm 0.13$  &  $43.57 \pm^{0.05}_{0.06}$ & $<$1.0    & 20.6/36 \\
PKS1607+26  & $<$2                   & $0.4 \pm 0.3$ & $4.2\pm^{1.5}_{0.6}$      &$44.27 \pm 0.01$  & $<$0.7  & 32.2/21\\
PKS2127+04  & $<19$ & 1.98 $^{+0.5}_{-0.4}$ & $0.49 \pm^{0.12}_{0.17}$  & $44.46 \pm 0.06$ & $<$0.9    & 14.7/16 \\

\hline
\end{tabular}
\end{center}
$^a$in erg~s$^{-1}$ \\
\end{footnotesize}
\end{table*}
%-------------- Tab.4
X-ray luminosities in this table and hereafter are K-corrected and
based on the best-fit model for each source.
Addition of a Gaussian profile to the baseline model
to characterise line emission never improved
significantly the quality of the
fit. The upper limits on the Equivalent Width
(EW) of an unresolved Fe $K_{\alpha}$ neutral fluorescent
line are also reported in Table~\ref{tab2}. They are generally
inconclusive.
Best-fit models
and residuals are shown in Fig.~\ref{fig:Spectra}.
%-------------- Fig.8
%------------------------- Fig.8
      
\begin{figure*}
\begin{center}
\includegraphics[height=60mm]{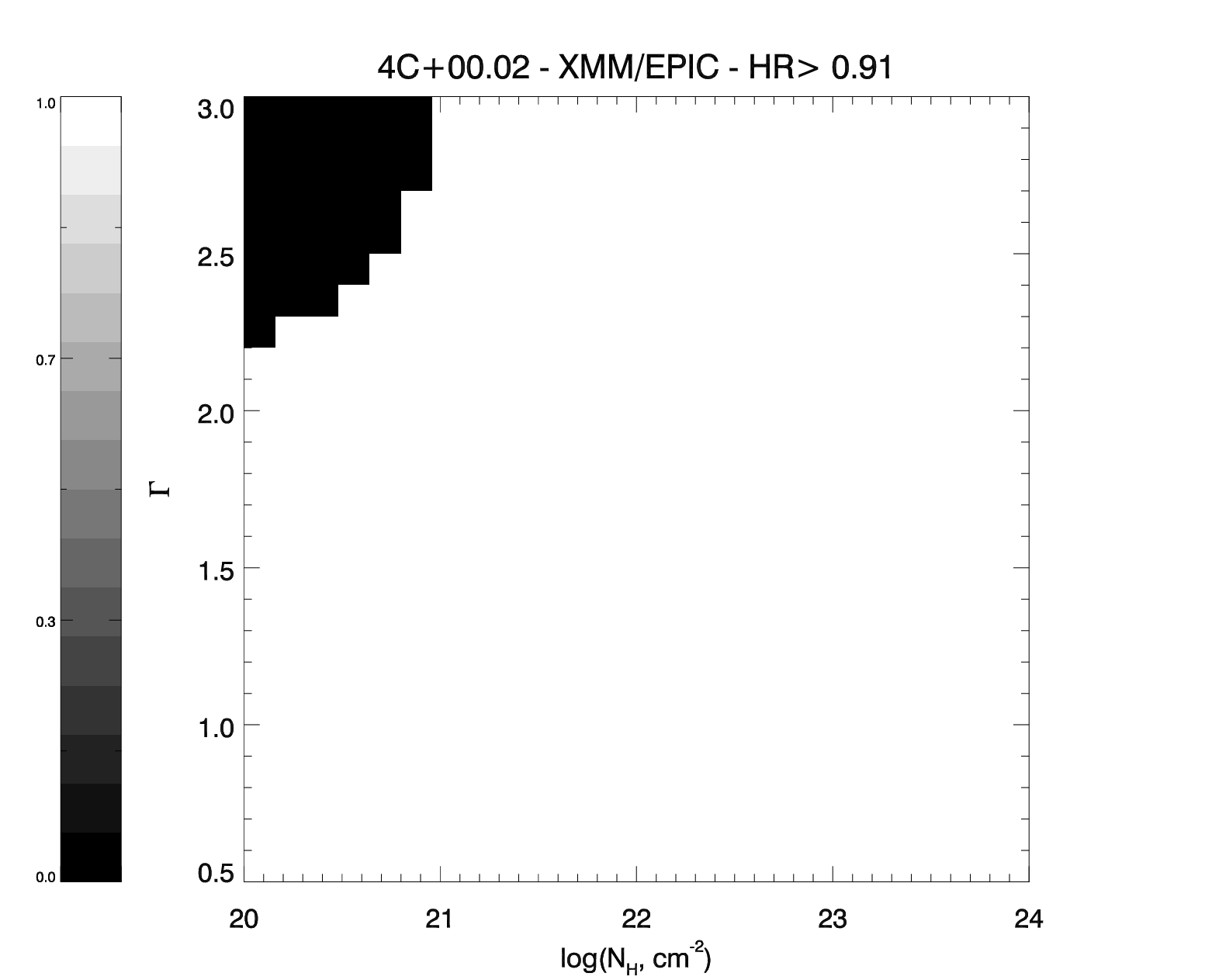}
\includegraphics[height=60mm]{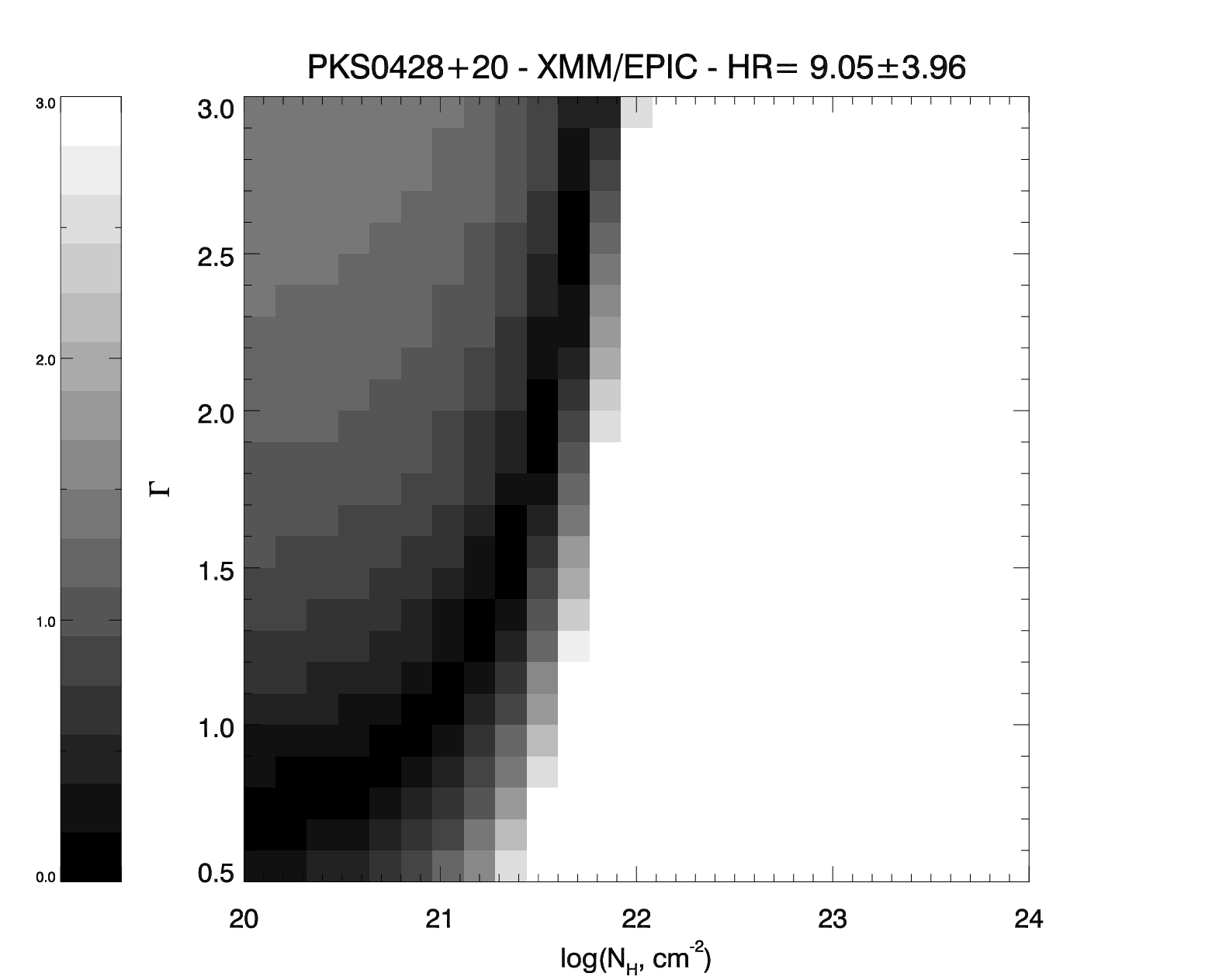}\\
\includegraphics[height=60mm]{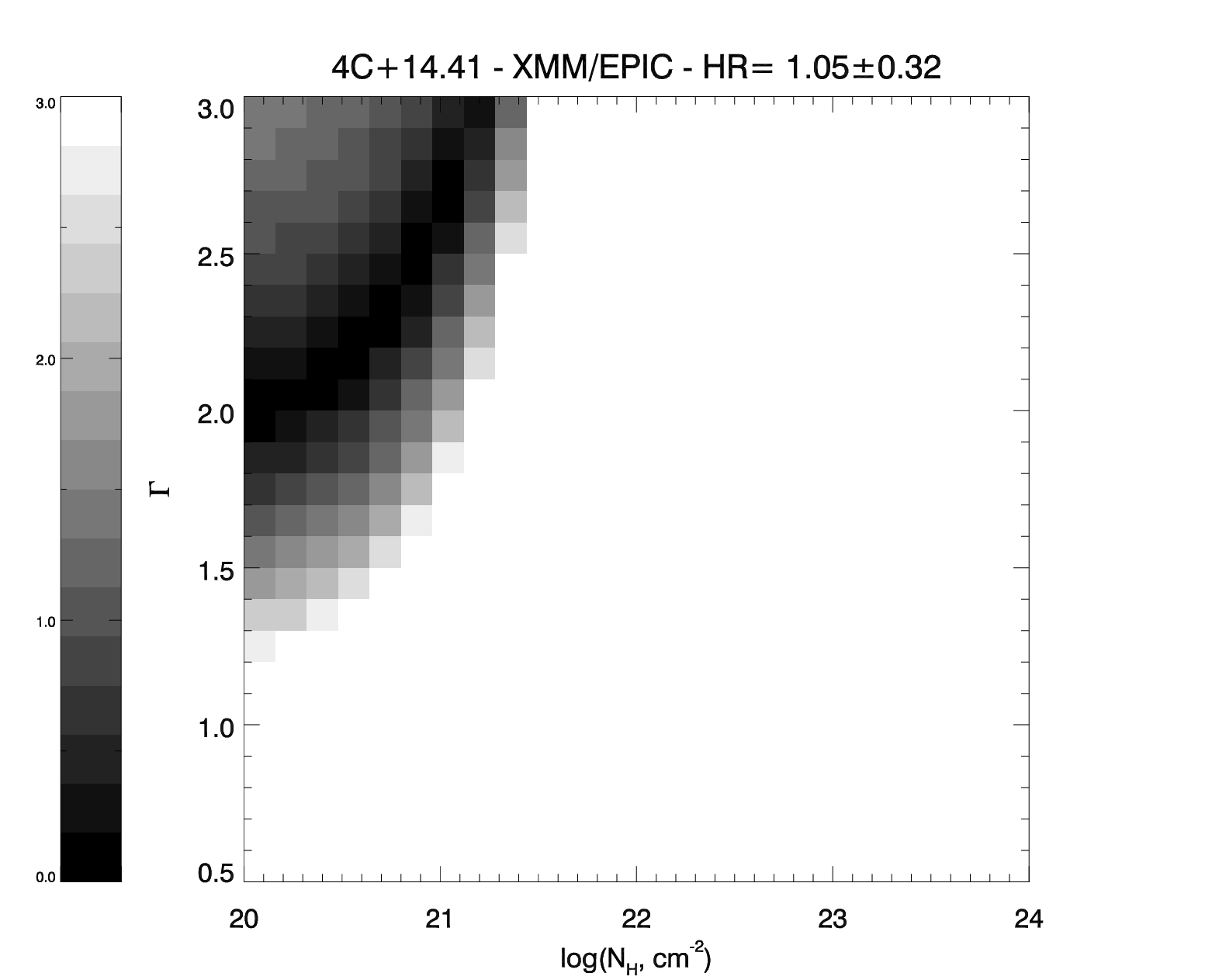}
\includegraphics[height=60mm]{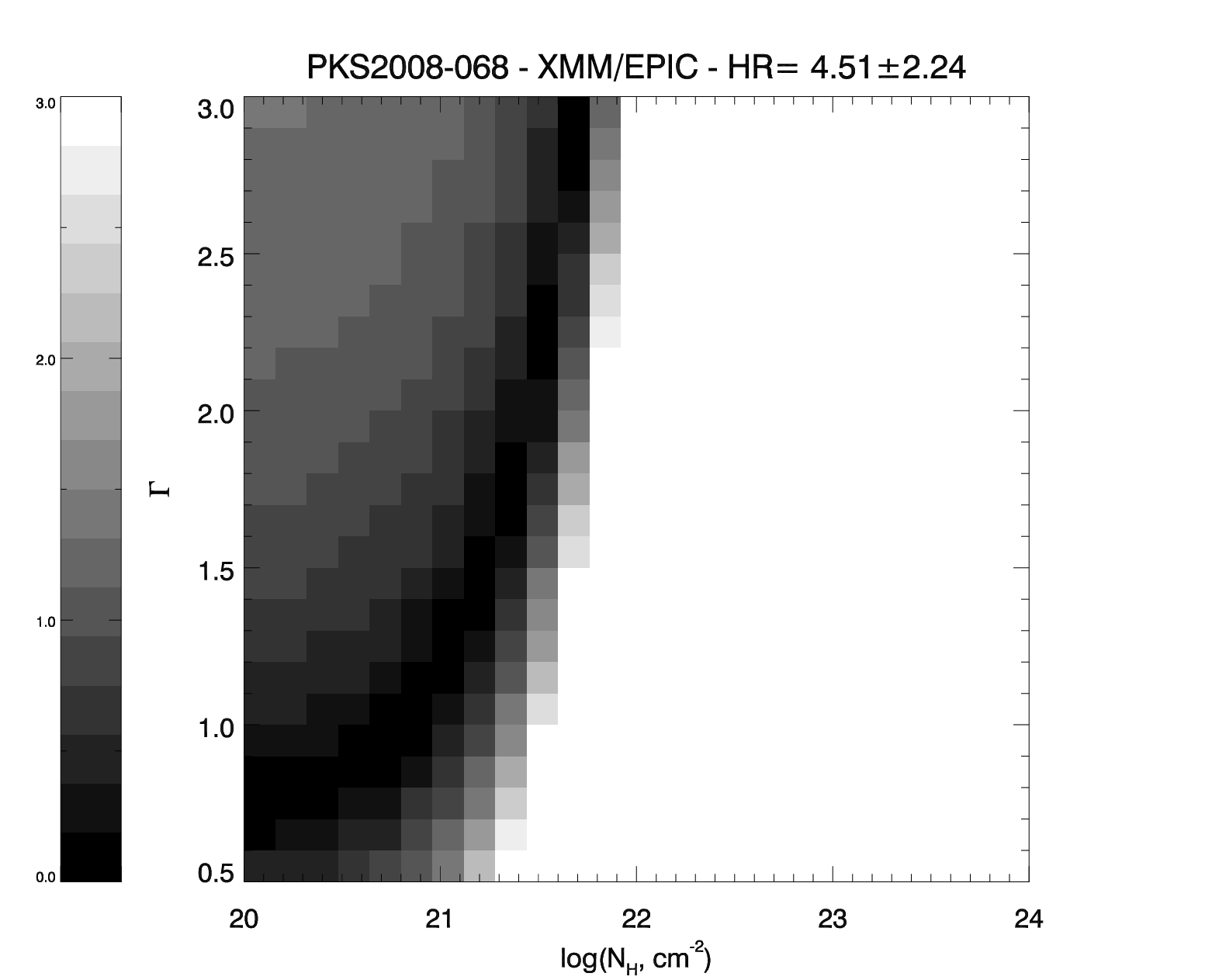}
\end{center}
\caption{Confidence level loci
in the ($\Gamma$,$N_H$)-plane. For 4C+00.02 the white
area shows the locus of the
parameter space where the model is consistent with the measured HR. In the other
panels the grey scale indicates the deviation from the measured HR
in units of standard deviations.}
\label{fig:contour}
\end{figure*}

For 4 other sources (4C+00.02, PKS~0428+20, 4C+14.41, and PKS~2008$-$068)
the low signal-to-noise did not allow for spectral analysis. We have therefore
based our estimation of the spectral parameters of the baseline
model on the Hardness Ratio (HR), here
defined as the ratio between the counts in the energy bands 1--10~keV and
0.2--1~keV. The measured HRs (or lower limit thereof) have been compared
with the predictions of grids of simulated baseline models.
Iso-HR contour plots in the $\Gamma$ ([0.5:3])
versus $N_H$ ([10$^{20}$,10$^{24}$~cm$^{-2}$]) parameter space
were built
(see Fig.~\ref{fig:contour})
%----------------- Fig.9
%----------------- Fig.9
The confidence interval in column density
was estimated as the minimum and maximum
value of the 1.6$\sigma$ interval around the curve corresponding to the
nominal HR, when 
$\Gamma$ was constrained in the range: [0.63,2.62].
The photon-index range corresponds to the $\langle \Gamma \rangle \pm
3 \sigma$ interval of the $\Gamma$ distribution, calculated on the
whole sample of GPS galaxy for which the data quality
allowed us to perform a
spectral analysis (cf. Fig.~\ref{fig1}).
The resulting constraints on the
column density are shown in Table~\ref{tab:Contour_nH}.
%----------------- Tab.3
\begin{table}
\begin{small}
\caption{\small Constraints on the intrinsic column density
derived from iso-HR contours in the $\Gamma$ versus $N_H$
parameter space. No constraint can be derived for 4C+00.02.}
\label{tab:Contour_nH}
\end{small}
\begin{footnotesize}
\begin{center}
\begin{tabular}{lccc} \hline \hline
Source         & $N_{H,max}$ & $F_{2-10}$$^a$ & $\log (L_{2-10})$ \\ 
&  (10$^{21}$~cm$^{-2}$)  & \\ \hline
PKS0428+20   & $<$6.9 & [0.3, 0.6] & 43.1 \\
4C+14.41     & $<$1.6 & [0.2, 0.5] & 43.1 \\
PKS2008$-$068 & $<$4.8 & [0.2, 0.5] & 43.8 \\
\hline
\end{tabular}
\end{center}
\noindent
$^a$ranges  in units of
10$^{-13}$~erg~s$^{-1}$~cm$^{-2}$
corresponding to the $N_H$ range as in this table, and to the
photon index in the range: [0.63,2.62] (more details in the text).
\end{footnotesize}
\end{table}
%----------------- Tab.3

\subsection{A Compton-thick AGN in PKS1607+26?}

\begin{figure}
\begin{center}
\includegraphics[height=80mm,angle=-90]{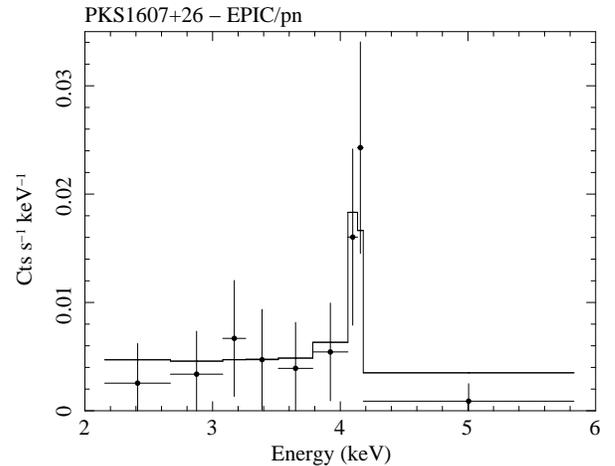}
\end{center}
\caption{EPIC-pn spectrum of PKS1607+26 in the 2--10~keV
energy band (observer's frame). The data points have been
rebinned such that each displayed spectral channel has a
signal-to-noise ratio larger than 1. The {\it solid line}
represents the best-fit reflection-dominated model (details in
text).
}
\label{fig6}
\end{figure}

\begin{figure}
\begin{center}
\includegraphics[height=80mm,angle=-90]{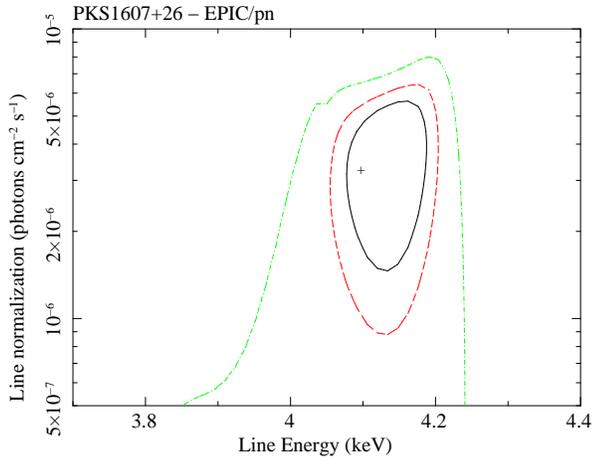}
\end{center}
\caption{Iso-$\chi^2$ contours in the centroid energy (observer's frame)
versus normalisation plane for the Gaussian profile in the
best-fit reflection-dominated model for PKS1607+26. The contours
correspond to the 68\%, 90\% and 99\% confidence level for two interesting
parameters, respectively.}
\label{fig8}
\end{figure}

\begin{figure}
\begin{center}
\includegraphics[height=80mm,angle=-90]{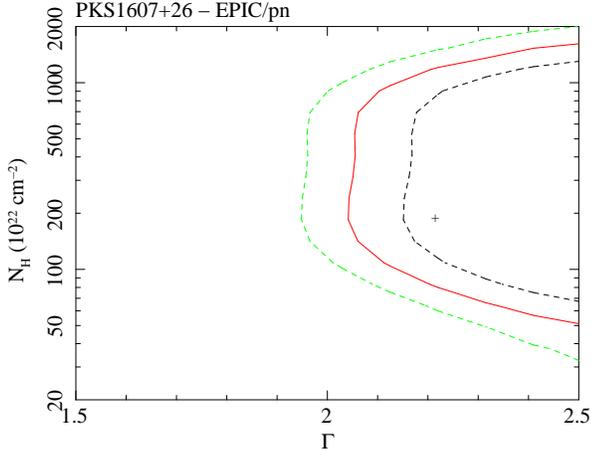}
\end{center}
\caption{Iso-$\chi^2$ contours for the column density covering
a power-law primary continuum derived from the EPIC pn
spectrum of PKS1607+26 in the reflection-dominated
scenario. The {\it lines} correspond to the 68\%, 90\%
confidence level for
one interesting parameter and to the 90\%
confidence level for two interesting
parameters, respectively.}
\label{fig10}
\end{figure}

The fit of the EPIC spectra of PKS1607+26 yields an unusually flat spectral
index: $\Gamma = 0.4 \pm 0.3$.
A flat spectrum may be indicative of a blazar-type spectral component.
Alternatively,
in radio-quiet AGN
spectral indices $\Gamma \simeq 0$ are generally interpreted
as evidence for a Compton-thick AGN, whose primary X-ray emission is
totally suppressed by optically thick matter with
a column density
$N_H > \sigma_t^{-1} \simeq 1.6 \times 10^{24}$~cm$^{-2}$
(see Comastri 2004 for a review). In Compton-thick AGN,
residual X-ray emission
red-wards the photoelectric cutoff
could be due to Compton-reflection of the
otherwise invisible primary radiation off the
obscuring matter\footnote{the GPS galaxy OQ+208 was the first
radio-loud Compton-thick AGN discovered (\cite{guainazzi04})}.
Fe K$_{\alpha}$ fluorescent emission
with large EW ($\approxgt 600$~eV; Risaliti 2002, Guainazzi et al. 2005)
is the unmistakable sign
of a Compton-thick AGN. 
We have therefore applied the baseline spectral model
for Compton-thick AGN to the PKS1627+06 pn spectrum
(the MOS spectra have very poor statistics, and do not provide
any further constraints):
a pure Compton-reflection continuum (model {\tt pexrav} in {\sc Xspec};
Magdziarz \& Zdziarski 1995) plus a Gaussian unresolved emission line.
In order to take advantage of the
full energy resolution of the EPIC cameras we fit the unbinned spectrum
using the Cash statistic, $C$ (\cite{cash76}). Although the Cash statistic
does not provide an absolute level of statistical confidence on
the quality of the fit, the resulting value
($C$=25.2/21~degrees of freedom) and the smoothness of the residuals
(Fig.~\ref{fig6}) indicate that the fit is good.
Once this continuum is adopted, an emission line is
%----------------- Fig.6
%----------------- Fig.6
required at a confidence level larger than 90\% for two interesting
parameters (or 95\% for one parameter; Fig.~\ref{fig8}).
%----------------- Fig.8
%----------------- Fig.8
The EW ($500 \pm 300$~eV) is consistent with values
typically observed in Compton-thick AGN.
However, the best-fit centroid energy of this feature
($E_c^{z=0} = 4.12 \pm^{0.06}_{0.05}$~keV) is
slightly inconsistent with Fe K$_{\alpha}$
fluorescence if converted into the rest frame with the quoted redshift
for this object: $E_c^{z=0.473} = 6.06 \pm^{0.09}_{0.07}$~keV. No
other known atomic transition could be responsible for a
feature at this energy.
We do not have a convincing explanation for this finding.
The redshift quoted in the literature for PKS1607+26
dates back to Philips \& Shaffer (1983).
Biretta et al. (1985) refer to previous mis-identification of this
object, and indeed the value of $z$=0.47 in the original paper
corresponds to the wrong identifier in Biretta et al. (1985).
If the emission line is indeed the Fe K$_{\alpha}$,
the redshift is $z \simeq$0.55.
Notwithstanding the ultimate redshift value, the very flat hard
X-ray spectral index, alongside
the detection of a strong emission-line features
point to
a Compton-thick nature for this object. In order to determine a lower-limit
to the column density covering the primary continuum, we have added to
the pure reflection model a power law obscured by photoelectric
absorption. We have used a combination of the models {\tt phabs}
and {\tt cabs} in {\sc Xspec} to take into account the contribution
of Compton scattering, which is no longer negligible for column densities
close to the Compton-thick limit. The obscuring column density is
constrained to be higher than $6 \times 10^{23}$~cm$^{-2}$ (90\%
confidence level for one interesting parameter; Fig.~\ref{fig10}).
%----------------- Fig.10
%----------------- Fig.10
This is the value we will associate with PKS1607+26
in all subsequents plots and distribution functions.
However, given the pending uncertainties on the ultimate identification
of this source, we will not use this measurement
to derive statistical estimators in the
$N_H$ distribution of the whole sample.

\section{Comparison with control samples of ``normal'' radio galaxies}

In this Section we present X-ray spectral properties of the
flux density and redshift-selected complete sub-sample
extracted from the Stanghellini et al. (1998) GPS sample
described in Sect.~2. Readers are
referred to the Guainazzi et al. (2006) and Vink et al. (2005)
for the spectral analysis of the sources not discussed in this paper.
We have repeated the analysis of the sources of the Vink et al. (2005)
with the same reduction and data screening criteria as in
Guainazzi et al. (2006) and in this paper. The results of our re-analysis
are consistent with theirs. 5~GHz luminosities are taken from Stanghellini
et al. (1998) and O'Dea (1998).

Our goal is also to
compare the properties of the complete radio-selected
flux-limited GPS
sample with a control sample of ``normal'' radio galaxies. The control
sample has been built from results recently published in the
literature. It is based on observations
of $z<1$ radio galaxies taken by ASCA (\cite{sambruna99}),
BeppoSAX (\cite{grandi06}), XMM-Newton and {\it Chandra} (FR~I:
\cite{donato04,evans06,balmaverde06}; FR~II
\cite{belsole06,evans06,hardcastle06}).
Only one measurement per source has been retained in the control sample,
based on the latest published result.
However, we have considered the latest {\it Chandra}
measurement, even when a later XMM-Newton observation
was available, under the assumption that the superior spatial resolution
of the {\it Chandra} optics provides a more reliable measurement of the
core emission.
The control sample comprises 93 sources (32 FR~I, 54 FR~II, the
remaining ones have no FR classification).
We stress that
the control sample is neither complete nor unbiased.
Moreover, it is not well matched with our GPS sample in redshift. The
probability that the redshift distribution of the GPS sample is the same
as in the entire control sample is 2\% only.
This difference is mainly due to FR~Is being generally at lower redshift
than our GPS sample, whereas our GPS and the control FR~II samples are
well matched in redshift (cf. Fig.~\ref{fig2}).
Whenever pertinent, we will
explicitly show in the following
the redshift dependence of the observables, and discuss any possible bias
associated with comparing samples inhomogeneous in redshift coverage.

\subsection{X-ray detection fraction}

We obtain a very large detection fraction;
all the sources of our sample yield a detection in
the soft X-ray band (0.5--2~keV), whereas 15 out of 16
are detected in the full band (0.5--10~keV). All of them
but one (PKS1345+125)
were unknown in X-rays prior to our {\it Chandra} (see also
Siemiginowska et al. 2008) and XMM-Newton observations.

\subsection{Spectral shape}

\begin{figure}
\centering 
\includegraphics[width=9cm]{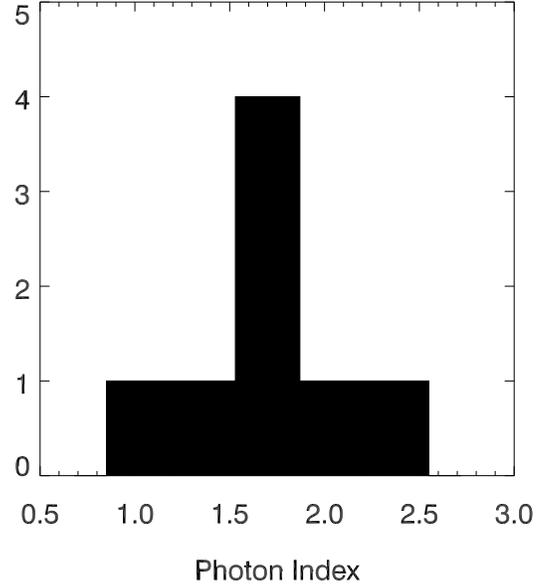}
\caption{
Distribution of the
photon index for the GPS sub-sample (8
objects), where data quality warranted spectral analysis.
}
\label{fig1}
\end{figure}

In Fig.~\ref{fig1} we show the distribution of spectral indices for
%------------------ Fig.3
%------------------ Fig.3
the 8 GPS of our sample\footnote{4C+32.44, 4C+62.22, B03710+439, COINSJ2355+4950, PKS0500+019, PKS1345+125, PKS2127+04, OQ+208},
where the number of counts is good enough for the
spectral analysis to be possible. The distribution has
a mean value $\langle \Gamma \rangle = 1.66 \pm 0.40$, and a standard deviation
$\sigma_{\Gamma} = 0.36$. The weighted mean is $\Gamma = 1.61 \pm 0.05$
if the measurements are weighted according to the inverse
square of their statistical uncertainties.

\subsection{Obscuration}

\begin{figure}
\centering
\includegraphics[width=9cm]{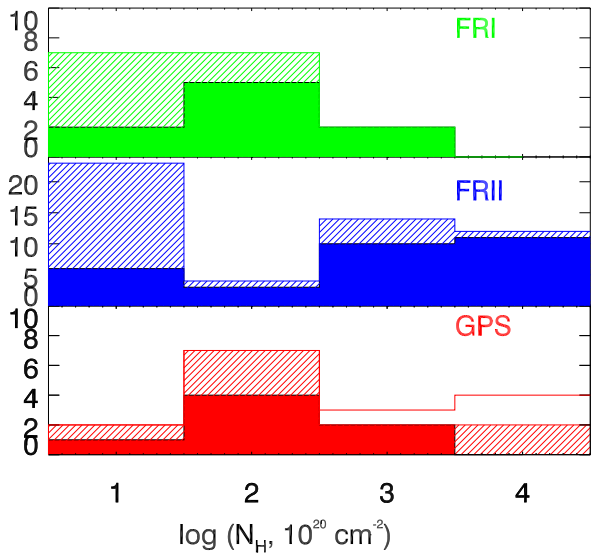}
\caption{
Distribution of the core obscuring column densities
for the GPS sample ({\it bottom panel}), the
FR~I ({\it upper panel}) and FR~II
 ({\it middle panel}) control samples.
{\it Shaded areas} indicate upper limits;
{\it empty areas} indicate lower limits.
}
\label{fig4}
\end{figure}

In Fig.~\ref{fig4} the distribution of column density for the GPS sources
of our sample is shown. The mean value of the distribution is
$\langle N_H^{{\rm GPS}} \rangle = 3 \times 10^{22}$~cm$^{-2}$ with a standard
deviation $\sigma_{N_H}$$\simeq$0.5~dex.
The same Figure shows
the comparison with the control sample.
%------------------ Fig.4
%------------------ Fig.4
The core emission of FR~I radio galaxies is generally unobscured or
only mildly obscured. In the Donato et al. (2004) FR~I sample, less than
one-third of the sample exhibits excess obscuration above the
Galactic contribution, with rest-frame column densities in the
range $10^{20-21}$~cm$^{-2}$, thus significantly lower than
observed in our sample.
The average of the column density distribution in the FR~I
control sample is
$\langle N_H^{{\rm FRI}} \rangle = (3.3 \pm
^{2.1}_{0.7}) \times 10^{21}$~cm$^{-2}$.

FR~II cores tend instead to include a
heavily obscured component. However, a detailed quantitative
comparison between the GPS and the FR~II control sample
is difficult.
For 11 out of 54
FR~II sources neither a measurement nor an upper limit on
the column density is available in the literature. These sources
are generally considered as unobscured (\cite{hardcastle06}).
Lack of inclusion of these sources could
potentially bias the control sample $N_H$ distribution
towards higher values. Bearing this caveat in mind, GPS galaxies
seems to fill a gap in the $N_H$ distribution between highly obscured
($\approxgt$$10^{23}$~cm$^{-2}$) and unobscured
($\approxlt$$10^{22}$~cm$^{-2}$) FR~II spectral components.

A  potential area of concern is the comparison
of column density measurements in samples, which are not well
matched in redshift.
%------------------ Fig.3
\begin{figure}
\centering
\includegraphics[width=9cm]{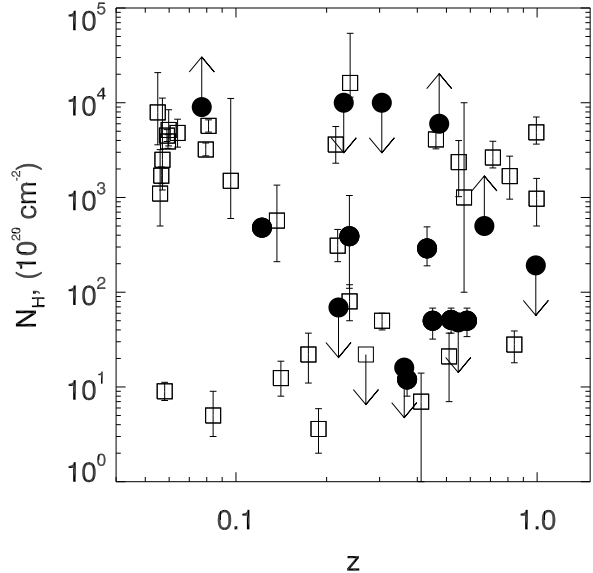}
\caption{
Core obscuring column density as a function of redshift for
the GPS galaxies ({\it filled dots}) and the FR~II control
sample ({\it empty squares}).
}
\label{fig2}
\end{figure}
%------------------ Fig.3
However, Fig.~\ref{fig2} shows that this bias is not
responsible for the difference between the average
of the column density distribution in the GPS and FR~II samples.
Moreover, low-redshift GPS galaxies exhibit
column densities not systematically
lower than high-redshift ones.

Hardcastle et al. (2006) remark that heavily absorbed nuclei are
rather common in narrow-line radio galaxies, whereas they are
comparatively rare in Low-Excitation Radio Galaxies
(LERG; \cite{laing94}). There are
7 LERGs in our control sample; 5 of them have no column density
measurement; the remaining two have column density of
$\simeq$$3 \times 10^{22}$~cm$^{-2}$ (3C~123; \cite{hardcastle06})
and $\simeq$$10^{23}$~cm$^{-2}$ (3C~427.1; \cite{belsole06}).
Taking into account the low number statistics and the uncertainties
on the column density upper limits on formally ``unobscured'' LERGs,
the comparison between X-ray obscuration in LERGs and GPSs
is inconclusive.

\begin{figure}
\centering
\includegraphics[width=9cm]{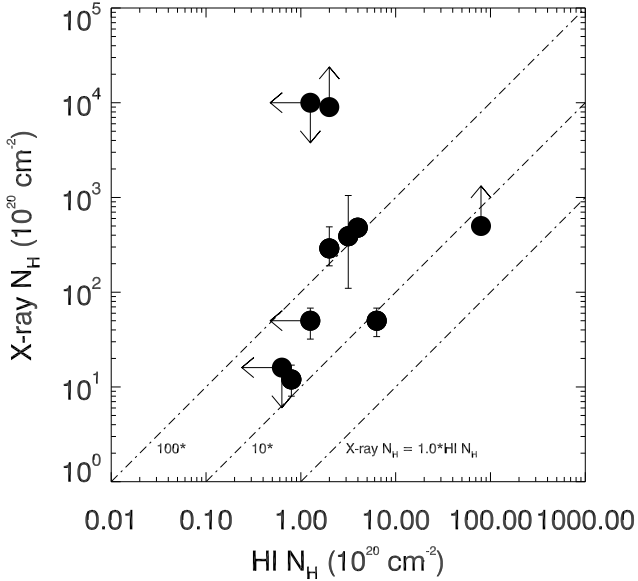}
\caption{
Comparison between the column densities measured in
X-rays ({\it this paper}) and with atomic hydrogen
observations (\cite{pihlstrom03}). The
{\it lines} represent loci of
constant X-ray versus HI column density ratio for
ratio values of 1, 10, and 100 (from {\it right} to
{\it left}), respectively.
}
\label{fig7}
\end{figure}

The X-ray column density is significantly larger than
the column density measured by HI observations
by a factor 10 to 100. The comparison is shown in Fig.~\ref{fig7}.
%------------------ Fig.7
%------------------ Fig.3
The estimate of the HI column density depends
on the values assumed for the spin temperature, $T_s$,
and for the fraction of background source covered
by the absorber, $f_c$. The data in Fig.~\ref{fig7}
assume $T_s$=100~K and $f_c$=1
(\cite{vermeulen03,pihlstrom03,gupta06b}).
The X-ray versus HI column density relation can be fit
with a zero-offset linear function if $T_s$ is of
the order of a few thousands K (\cite{ostorero09}).
Alternatively, a low covering fraction could be
responsible for the large X-ray to HI column density
ratio, although this explanation is less likely
given the large column densities measured in X-rays.

\begin{figure}
   \centering
\hbox{  
   \includegraphics[width=9cm]{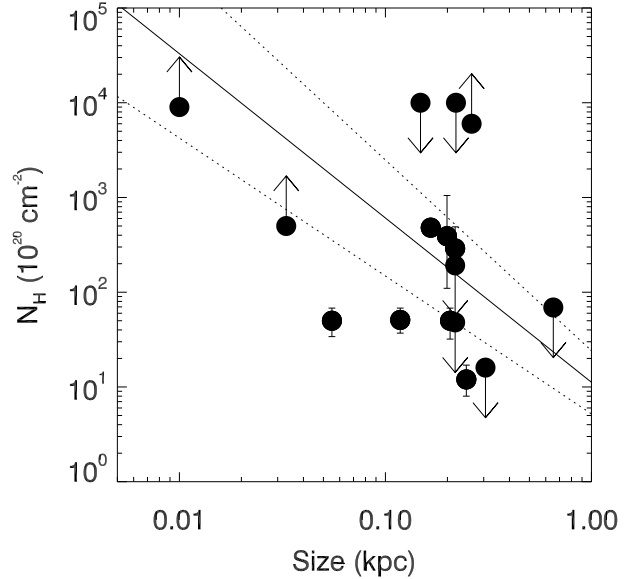}
}
\caption{X-ray column density versus the size
of the radio structure. The {\it solid} line
represent the best censored data linear fit
with a function: $\log(N_H) =
A + B \times \log(r_{kpc})$; the {\it
dotted lines} correspond
to $\pm 1 \sigma$ uncertainties on the best-fit
parameters
}
\label{fig3}
\end{figure}

Holt et al. (2003) proposed an ``onion-skin'' model for
the nuclear environment gas in B1345+125, to explain the
reddening properties of the different components
of the optical lines. The jet would
pierce its way through a dense
cocoon of gas and dust, with decreasing density at larger
distances from the radio core. If this scenario would
apply to the whole class of GPS samples, and
X-ray and radio emission originate in the same
physical system, one
might expect an anti-correlation between the measured
column density and the size of the radio structure, with
a large scatter due to the unknown line-of-sight angles
distribution. This correlation
is shown in Fig.~\ref{fig3}. 
%------------------ Fig.4
%------------------ Fig.4
A censored fit on this data with a function: $\log(N_H) =
A + B \times \log(r_{kpc})$ yields: $A = 21.4 \pm 0.4$, and
$B = -1.2 \pm 0.3$, where $r_{kpc}$ is the size
of the radio structure in
units of kpc. This result is formally robust,
but admittedly mainly driven
by the two extreme data points. A confirmation of this
correlation by increasing the number of small ($r <$100~pc)
and large ($r > 1$~kpc) objects
for which spectroscopic X-ray data are
available is a task we are actively pursuing.
Interestingly enough, an anti-correlation
between the linear dimension of the sub-galactic radio galaxy and
the radio HI column density was discovered by
Pihilst\"om et al. (2003). 
As already pointed out by Gupta et al. (2006), this
anti-correlation could be driven by small sources probing
gas closer to the AGN and hence at a higher spin
temperatures.
In this context, it is also interesting to observe that
GPS {\it quasars} exhibit no absorption (with upper limits
$\sim 10^{21}$~cm$^{-2}$; \cite{siemiginowska08}), as well
as diffuse emission associated with jets, binary structures or
embedding clusters.
The detection rate as well as the column density of HI
absorption increases with core prominence (\cite{gupta06a}).
The core prominence is a statistical indicator of the
orientation of the jet axis to the line of sight.
On average HI absorbers are more common
and exhibit higher column densities in galaxies than
in quasars. This can be explained if the HI
absorbing gas is distributed in a circumnuclear disk
much smaller than the size of the radio emitting
region, and only a small fraction of it is obscured in
objects at large inclinations.

\subsection{Radio-to-X-ray correlations}

\begin{figure*}
   \centering
\hbox{  
   \includegraphics[width=9cm]{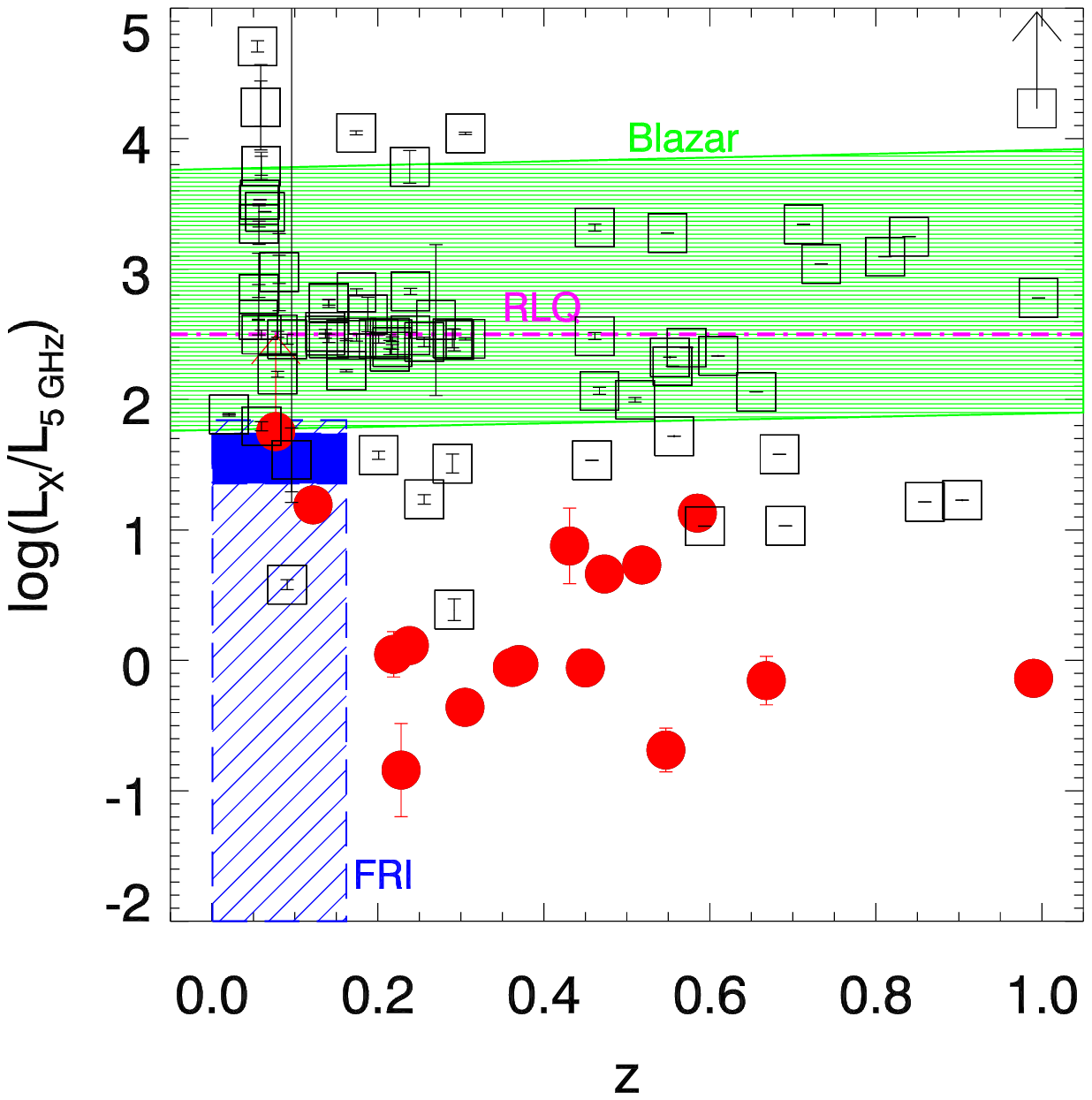}
   \includegraphics[width=9cm]{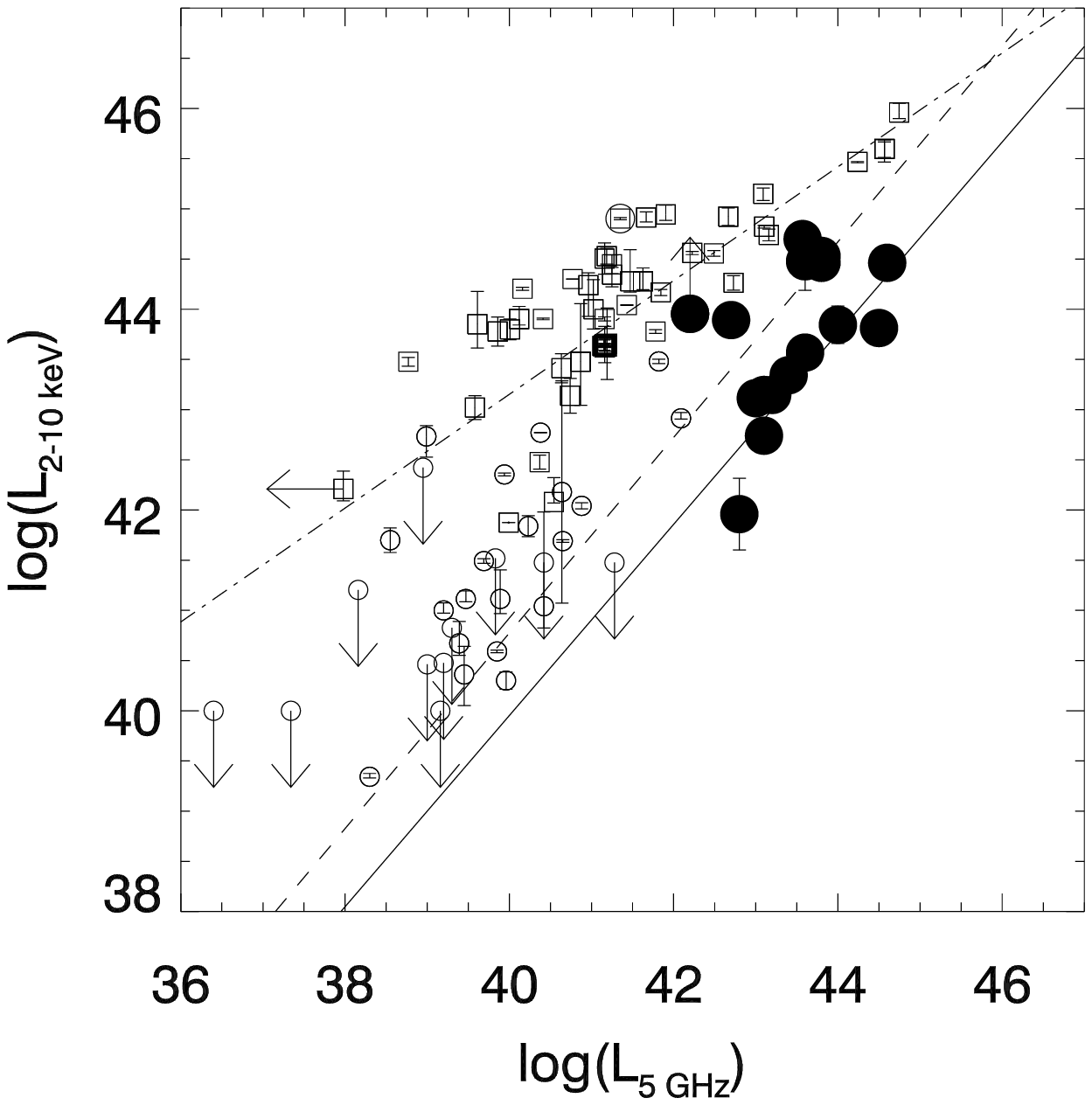}
}
\caption{
{\it Left panel}:
2--10~keV versus 5~GHz logarithmic ratio versus
redshift for the GPS ({\it filled dots})
and the FR~II sample ({\it empty squares}). The
{\it obliquely shaded box} indicates the locus
of the FR~I {\it Chandra} sample;
the {\it horizontally shaded box}
the locus of the blazar sample of Fossati et al.
(1998); the {\it dot-dashed} line the locus
corresponding to a typical Spectral Energy
Distribution of a radio-loud quasar after
Elvis et al. (1994b). 
{\it Right panel}: 2--10~keV versus
5~GHz core luminosity for the GPS galaxies
({\it filled circles}), and a control
sample of radio galaxies (FR~I:{\it empty circles},
FR~II: {\it empty squares}).
The {\it
lines} represent the best-fit
regression line for censored data for
X-ray weak GPS ({\it continuous}), FR~I
({\it dashed}) and FR~II
({\it dashed-dotted})
}
\label{fig5}
\end{figure*}

In the {\it left panel} of
Fig.~\ref{fig5} we compare the logarithmic ratio between the
%------------------ Fig.4
%------------------ Fig.4
2--10~keV and the core 5~GHz luminosity
($L_{5GHz} \equiv \nu_{5GHz} \times S_{5GHz}$,
where $S_{5GHz}$ is the luminosity density at
5~GHz) for the
GPS and the control sample.
Values for the GPS sample
range between -0.5 and 1.5. No clear dependence on
redshift is observed.
GPS galaxies are X-ray under-luminous by about an
order of magnitude with respect to
their radio power once compared to FR~II radio galaxies,
blazars (\cite{fossati98})
and radio-loud quasars (\cite{elvis94b}).
On the other hand, the X-ray-to-radio luminosity ratios
in GPS galaxies well match
values observed in FR~I galaxies:
$\log (L_X/L_{5GHz}) < 1.8$. It is important to
stress again that there is almost no overlap in redshift
between the GPS and the FR~I samples, though.

The interpretation of the X-ray-to-radio luminosity
correlation depends on the origin of the bulk of the
VLA radio emission in compact galaxies. VLBI
observations of GPS galaxies revealed a fraction
of Compact Symmetric Objects (CSO) between $\simeq$30\%
and 100\% (Stanghellini et al. 1997, 1999; \cite{xiang05,liu07}).
Three objects in our sample show a CSO
morphology: PKS~0050+019, PKS~1345+125, and PKS~2008$-$068
(Stanghellini et al. 1997, 1999), although in all
these cases the morphology is rather complex, with
multiple components on scales $\approxlt$20~pc.
High-resolution, multi-frequency observations of
our sample would be required to
estimate which fraction of the VLA flux
can be safely attributed to a core.

From Fig.~\ref{fig5} a possible bimodality of the
X-ray-to-radio luminosity ratio in the GPS sample is apparent.
The fit of the
cumulative distribution function of this quantity with a single
Gaussian yields a Kolmogornov/Smirnov value of 0.33,
corresponding to null hypothesis probability of about
4\%. A fit with a double Gaussian yields instead a value of
0.18, with a null hypothesis probability of 65\%.
We consider this as a tentative piece of evidence for
bimodality only.
We will refer in the following to ``X-ray bright'' and
to ``X-ray weak'' GPS galaxies as those, whose
$\log (L_X)/(L_{5~GHz})$ ratio is larger/smaller than 0.5,
respectively.
No significant difference in the spectral shape
between the the ``X-ray bright'' and ``X-ray weak'' sources was
observed. In particular, the obscuring column density
distributions are indistinguishable.

In the {\it right panel} of
Fig.~\ref{fig5} the 2-10~keV luminosity
is plotted against the 5~GHz luminosity. In FR~I galaxies
a strong correlation between core X-ray, radio and
optical flux is known (\cite{chiaberge99,hardcastle00}).
In our control sample, the slope of this correlation
is consistent with unity: $0.97 \pm 0.06$.
Interestingly enough, this slope is consistent
with the slope observed in ``X-ray weak'' GPS galaxies
$0.95 \pm 0.12$,with a $\simeq$0.5~dex
offset at face-value. The ``X-ray bright'' sample
exhibits a significantly flatter slope, although
with large uncertainties ($0.1 \pm 0.2$).
Again, it is hard to estimate any bias associated with
the coarse spatial resolution of the GPS radio measurements.

\section{Discussion}

In this Section we will review the main
X-ray observational
properties of our whole GPS sample, trying to address the
three issues which originally motivated our study:

\begin{itemize}

\item What is the physical origin of the X-ray emission
in GPS galaxies?

\item Which physical system is associated with the X-ray obscuration?

\item What is the ``endpoint'' of the evolution of
compact radio sources?

\end{itemize}

\subsection{X-ray spectral support for an accretion-disk origin}

The distribution of spectral indices in the GPS sample
by itself does
not provide any stringent constraints
on the origin of X-ray emission
in GPS galaxies. The
mean value, $\langle \Gamma \rangle = 1.66$
($\sigma_{\Gamma} = 0.36$), is consistent
with spectral components associated with the jet
in radio galaxies
($\Gamma = 1.88 \pm 0.02$), as well as with
accretion ($\Gamma = 1.76 \pm 0.02$;
Evans et al. 2006), although is nominally closer to the latter.
A possible clue may come from the fact that
the column density measured in X-rays is invariably
larger by 1-2 orders of magnitudes than that measured
in radio. This finding could be naturally explained
by X-rays being produced in a smaller region than the radio.
On average, the radio morphology of compact radio sources
strikingly resembles that of large-scale radio doubles,
although on a scale which is entirely confined within
the optical narrow-line emission regions.
Radio emission traces therefore
the radio hotspot and lobes. X-rays could be instead
generated by the base of the jet, or
by the accretion disk.

Support for the X-rays arising from a relatively compact region
comes from the comparison
with  radio-quiet AGN. Once a similar baseline
model is employed, Seyfert galaxies have:
$\langle \Gamma_{Sy} \rangle = 1.64 \pm 0.05$
(Bianchi et al., submitted)\footnote{This is {\it
not} the intrinsic spectral index, which
can be significantly larger due to the hardening
effect of ionised absorbers of disk/torus reflection.}.
GPS galaxy X-ray spectra
lack the complexity that Seyferts typically exhibit. There
is no strong evidence for a soft excess (with the
only exception of OQ+208; Guainazzi et al. 2004), warm absorber,
warm scattering or Fe K$_{\alpha}$ fluorescent emission
(with, again, the notable exceptions of OQ+208 and,
possibly PKS1607+26) in our sample. However, most of the GPS galaxy
spectra collected so far with either XMM-Newton or
{\it Chandra} do not possess the statistical quality that
would be needed for these additional spectral features to
be unambiguously detected.

The scatter in the X-ray to radio luminosity ratio, and its lack
of dependence on source size and age (cf. also Fig.~\ref{fig12})
may indicate a link between accretion disk and jet activity
primarily driven by disk instabilities. 10-20\% of GPS objects
exhibit very
extended radio (\cite{baum90,stanghellini90,schoenmakers99,marecki03})
or X-ray (\cite{siemiginowska02}, 2003) emission. These components have
been interpreted as remnants of past enhanced activity.
A similar behaviour on much shorter time scales is observed
in Galactic Black Hole Candidates and micro-quasars, such
as GRS~1915+105 (\cite{belloni00,fender04}). Jet blobs are
supposed to be fed
by the evacuation of the innermost accretion disk regions.
This mechanism
yields alternating X-rays- (disk-dominated) and radio-bright
(jet blob-dominated) phases. Models based on
disk radiation pressure instabilities reproduce well the
timescales of these transitions (Czerny et al., in preparation),
although they do not make yet specific predictions
on the evolution of the spectral energy distribution.

\subsection{Dilution by X-rays from radio-emitting regions?}

Most likely, the baseline model is indeed too simple.
X-ray absorption could be ``diluted'' by X-rays coming
from the high surface-brightness radio components,
thus complicating the interpretation of the results
derived by our simple baseline model. This scenario would
also explain
the (still tentative) anti-correlation between the 
X-ray column density and the size of the radio source
(Fig.~\ref{fig3}).
In larger sources, a larger fraction of the X-ray emission
associated with the radio hot-spots or lobes may be
visible beyond the rim of the obscuring matter. Should
this indeed be the case, we should, however, observe deviations
from a simple power-law spectral shape. We
indeed observe a soft excess in the radio-loud
Compton-thick GPS galaxy OQ+208. This is the closest
object in our sample, and the only one where high-resolution
spatially-resolved spectroscopy with {\it Chandra} could provide
direct observational clues on the
physical location of the X-ray emitting plasma.

\subsection{The X-ray obscuring environment}

\begin{figure}
   \centering
\hbox{  
   \includegraphics[width=9cm]{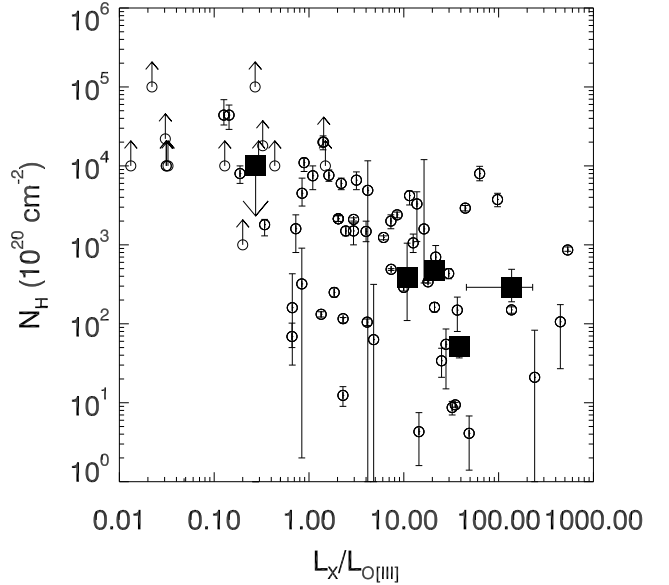}
}
\caption{X-ray column density versus the
ratio between the absorption-corrected 2--10~keV
and the {\sc O[iii]} fluxes. The {\it filled
squares} are the 5 GPS galaxies in our
sample for which {\sc O[iii]} measurements
are available; the {\it empty circles}
represent a sample of Seyfert~2 galaxies
after Guainazzi et al. (2005).
}
\label{fig11}
\end{figure}

Lacking other direct evidence from X-rays
alone, one may use multiwavelength diagnostics to obtain
further clues on the origin of the X-ray emission in
GPS galaxies. A well established diagnostic tool for
X-ray obscuration in radio-quiet AGN involves the
comparison between the column density measured in X-rays
and the absorption-corrected ratio of
X-ray to {\sc O[iii]} fluxes (\cite{maiolino98}).
In the context of Seyfert galaxies, this
diagram is used to calibrate the latter quantity as
an estimator for obscuration. We use here the same plot with
a different purpose, namely to test whether the ionising continuum
powering the Narrow Line Regions in GPS galaxies has
the same properties as in Seyfert galaxies once
normalised to the X-ray primary emission. If this is
the case, one may conclude that the ``normalising
primary continua'' share the same properties
in the two populations. The results of this comparison
are shown in Fig.~\ref{fig11}. 5 objects in our GPS sample
%------------------ Fig.11
%------------------ Fig.11
have spectroscopic {\sc O[iii]}
measurements (O'Dea 1998, Labiano et al. 2005, 2008
and references therein).
The ``control sample'' is a collection of
Seyfert~2-galaxy measurements in Guainazzi et al. (2005).
The agreement is good. Although more GPS data would be
required to ensure a homogeneous coverage of this plane,
this evidence further points towards an accretion
origin for the X-ray emission in GPS galaxies.

Interestingly, the fraction of Compton-thick AGN in the
GPS sample ($0.12 \pm 0.09$, if PKS1607+26 is
considered) is comparable to fractions observed
in radio-quiet AGN (\cite{heckman05}).
Conversely, only one Compton-thick
AGN has been detected in large-scale radio
galaxies (\cite{erlund08}), although FR~II typically exhibit
spectral components with very large obscuration
(\cite{belsole06,evans06}).

\subsection{X-ray emission entirely associated with radio-emitting regions?}

We now investigate the implications of the difference between X-ray
and radio {\sc HI} column densities, possibly due
to the ionisation state of a single gas system
simultaneously covering the radio and the X-ray
emission.
Attributing this difference solely to ionisation effects would
mean ionisation fractions of 90\% to 99\% (see the discussion
in Vink et al. 2005). Photoionisation simulations with {\sc Cloudy}
(\cite{ferland98})
show that this corresponds to
an ionisation parameter $\approxgt 20$
for a gaseous nebula photoionised by a typical AGN
continuum. 
An anti-correlation between the HI column density
and the linear projected size of the radio source
is now well established (\cite{vermeulen03,pihlstrom03,gupta06a}).
Smaller sources ($< 0.5$~kpc)
tend to have higher {\sc Hi} column density than
larger sources ($> 0.5$~kpc).
If not driven by uncertainties in the
spin temperature of the HI absorbing gas, this
anticorrelation can be explained by GPS galaxies hosting young
radio sources, which evolve in a disk distribution of
gas with a power-law radial density dependence. The same
explanation could lie behind the (tentative)
anti-correlation between X-ray column density
and radio size in GPS galaxies (Fig.~\ref{fig3}).
One can assume a scenario where the radio and
X-ray source are seen through the same line-of-sight
and are both embedded in a screen of ionised gas, responsible
for X-ray photoelectric absorption as well as for radio
free-free absorption. From the definition of the
ionisation parameter, it follows for the X-ray
regime:
$$
n_{e,3} \approxgt 10^2 R_{kpc}^{-2} L_{44}
$$
where $n_{e,3}$ is the plasma density in units of
$10^3$~cm$^{-3}$, $R_{kpc}$ is the distance between
the ionising source and the innermost side of the
absorbing material in kpc, and $L_{44}$ is the ionising
luminosity in units of $10^{44}$~erg~s$^{-1}$.
It is plausible that densities in GPS sources
are high enough that free-free absorption may
play a role (\cite{odea91,stawarz08}).
The emission measure for free-free absorption at
5~GHz is
(see, {\it e.g.}, Osterbrock 1977):
$$
n_e^2 d_{pc} \simeq 90 \tau T_4^{1.35}
$$
where $d_{pc}$ is the path length 
in parsecs and $T_4$ is the
temperature in units of $10^4$~K. Assuming
$R \sim d$, one can constrain the
plasma temperature:
$$
T_4 \approxgt (L_{44} \eta_{30})^{1.5} R_{kpc}^{-2.2}
$$
where $\eta_{30}$ is the conversion factor between the
absorption-corrected 2--10~keV  and the bolometric
luminosity divided by 30. For standard values of $\eta_{30}$
(\cite{elvis94b}), almost
all GPS sources in our sample require temperatures above
that at the dust sublimation radius in this scenario. 
This outcome could be tested by measurements of the
optical reddening in our sample.

If the compact jet ``drills'' its way through
such a medium, would it be significantly decelerated
or even ``frustrated'', i.e. permanently confined? This
issue was discussed by Guainazzi et al. (2004) for
the case of the Compton-thick absorber in OQ+208.
They concluded that for large inclination angles or
large thickness of the Compton-thick layer, the jet
could have been significantly decelerated by the interaction
with the ambient medium. Although permanent confinement
is unlikely in OQ+208, underestimating the evolution time scale
by one-two orders of magnitude is possible. We revise
their argument for the whole sample of GPS galaxies.
The expansion time for a jet propagating in
pressure equilibrium through in homogeneous medium can
be expressed as (\cite{scheuer74,carvalho85,carvalho98}):
$$
t_e \sim 6  \hspace{0.1cm} r^{1.5}_{100~pc} L_{inj,44}^{-0.5} (\cos \imath)^{-1.5} \Omega_{10}^{0.5} N_H^{0.5} \hspace{0.1cm} \hbox{s}
$$
where $L_{inj,44}$ is the (unknown) luminosity injected in the jet
in units of $10^{44}$~erg~s$^{-1}$, and $\Omega_{10}$ is the jet
opening angle in units of 10~degrees.
If we assume the censored best-fit for the
linear size versus X-ray column density relation:
$$
t_e \sim 4 \times 10^5  \hspace{0.1cm} r^{0.9}_{100~pc} L_{inj,44}^{-0.5} \Theta  \hspace{0.1cm} \hbox{years}
$$
where we have compacted all the geometrical factors in the
variable $\Theta \equiv (\cos \imath)^{-1.5} \Omega_{10}^{0.5}$.
There are admittedly several uncertainties in deriving the
above scaling law.
Most importantly, we know very little of
the actual gas distribution in the environment surrounding
compact radio sources (as well as in any other type of
AGN). Still, the above estimate indicates that a GPS
may expand during a time of the order of at least
$10^5$~years. This is longer
than currently existing observational estimates
(Poladitis \& Conway 2003; Gugliucci et al. 2005),
but not enough to make the hypothesis of permanent confinement
viable.

\subsection{Evolution of GPS sources}

If the jet can indeed survive its eventful youth, and grow
to reach a full level of kpc-scale maturity and beyond, what
would it look like? The radio-to-X-ray luminosity plane does
not ultimately elucidate the possible connection between GPS
and ``mature'' radio galaxies. GPS galaxies are intriguingly well
aligned along the
extrapolation at high radio power of the
correlation between radio core
and X-ray luminosity valid for FR~I radio galaxies (\cite{evans06}).
This correlation was proposed as evidence
supporting a jet
origin of unabsorbed X-ray spectral components in FR~I
radio galaxies. On the other hand, GPS galaxies have a comparable
X-ray luminosity to FR~IIs (cf. Fig.~\ref{fig5}). In FR~IIs,
the obscured X-ray spectral component is probably
associated with accretion onto the supermassive black hole
obscured by a ``torus'', in analogy to the accepted
paradigm applicable to radio-quiet AGN (\cite{piconcelli08}).
If the X-ray
emission in GPS galaxies is due to accretion, the
evolution of the radio and X-ray wavebands could be totally
decoupled. Radio power would decline with the linear size
(see, {\it e.g.}, Fanti et al. 1995) while the sources
expand through the ISM; at the same time the accretion disk
would maintain a stable flow. At the end of their infancy,
GPS galaxies would reach
maturity as FR~II radio galaxies.

\begin{figure}
   \centering
\hbox{  
  \vspace{-0.5cm}
   \includegraphics[width=9cm]{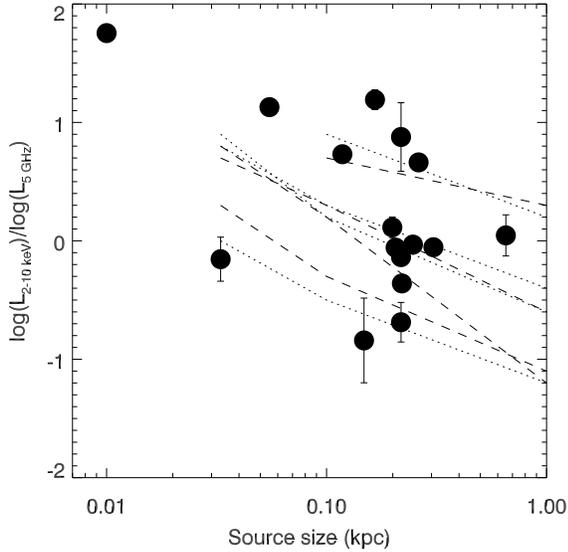}
}
\caption{X-ray to radio luminosity ratio
versus the linear size for the sources of our GPS sample.
The {\it lines} indicate the predictions of the
Stawarz et al. (2008) model for jet kinetic
powers ranging from $10^{44}$~erg~s$^{-1}$ to
$10^{47}$~erg~s$^{-1}$. {\it
Dashed lines}: power-law injection function (Fig.~2
in their paper);
{\it dotted lines}: broken power-law injection function (Fig.~3
in their paper).
}
\label{fig12}
\end{figure}

Evolutionary scenarios require that the radio luminosity
of GPS sources decreases with evolution, not
to exceed the number of observed FR~II galaxies
(\cite{readhead96}). Recently, Stawarz et al. (2008)
have proposed an evolutionary model for GPS sources,
which explicitly predicts the dependency of the
broadband Spectral Energy Distribution
on the source linear size. In their model
high-energy emission is produced by
upscattering of various photon fields
by the lobes' electrons. They predict a decrease of
the X-ray to radio luminosity ratio by 1--2 orders of
magnitude when the GPS source size increases
from 30~pc to 1~kpc. In Fig.~\ref{fig12}
%------------------ Fig.12
%------------------ Fig.12
we compare this prediction with our observations. There is
no evidence for a strong anti-correlation between the
two quantities; the slope of the best linear fit is
$0.3 \pm 1.5$ (1-$\sigma$ statistical error). Still,
the observational data are consistent with the Stawarz et al.
model predictions, if the GPS galaxies in our sample cover a wide
range in jet kinetic power.

It is still
impossible with
the available data to
decide between an evolutionary scenario in which the
source evolves
into a conventional FR~II or FR~I radio galaxy. To achieve this
goal, photometric X-ray observations of sizable samples of
low-luminosity ($L_{5~GHz} \simeq 10^{40-42}$~erg~s$^{-1}$)
GPS galaxies would be crucial. Fortunately, it is now
in principle possible to perform this experiment, thanks to
the point-like source sensitivity and scheduling flexibility
of {\it Chandra}.

\section{Summary and conclusions}

This paper
reports our current knowledge on
X-ray emission in GPS galaxies. For the first time a 
complete radio-selected sample of GPS galaxies
has been almost entirely
observed with a modern X-ray observatory (mostly
with XMM-Newton). The sample is
comprised all the $z < 1$ sources of the
Stanghellini et al. (1998) sample having a
5~GHz flux density
$\ge$1~Jy. The main
results of our study can be summarised as follows:

\begin{itemize}

\item we obtain a very large detection fraction;
all the sources of our sample yield a detection in
the soft X-ray band (0.5--2~keV), whereas 15 out of 16
are detected in the full band (0.5--10~keV)

\item in almost all cases, a simple power law
modified by photoelectric absorption represents
an adequate description of the 0.5--10~keV spectrum.
In a few objects
the number of net counts is
not good enough to allow a full spectral analysis. In this case
basic spectral parameters were derived from hardness ratios
assuming the baseline model above

\item the mean of the spectral indices distribution
is $\langle \Gamma \rangle = 1.66$
($\sigma_{\Gamma} = 0.36$). Although the uncertainty
on this parameter is still too large
to pinpoint the physical mechanism
responsible for the observed X-ray emission, at face
value the $\Gamma$ distribution is closer to
those AGN classes whose
X-ray emission is believed to be dominated by accretion:
Seyfert Galaxies and the obscured spectral component in
FR~II radio galaxies

\item we report the possible discovery of a Compton-thick
AGN in PKS1607+26.
Radio-loud Compton-thick AGN are still an
elusive population (\cite{comastri04}). Together
with OQ+208 - the other Compton-thick AGN in the sample -
PKS1607+26 is the only GPS galaxy where
an X-ray emission line has been detected,
possibly associated with Fe K$_{\alpha}$
fluorescence

\item X-ray spectra of GPS galaxies are significantly obscured.
The mean value of the column density distribution
(without PKS1607+26, due to pending uncertainties on
the identification of this source) is
$\langle N_H^{{\rm GPS}} \rangle = 3 \times 10^{22}$~cm$^{-2}$ with a standard
deviation $\sigma_{N_H}$$\simeq$0.5~dex.
Such a value is much larger than column densities measured in
a control sample of FR~I radio galaxies, but still
less than column densities covering accretion-related
X-ray spectral components in FR~II radio galaxies
(\cite{balmaverde06,belsole06,evans06}).

\item the X-ray column density measured in almost all GPS
galaxies is larger than the HI column density
measured in the radio by a factor of 10 to 100.
This could be the signature of
physically different absorption
systems (\cite{vink05,guainazzi06})
or of a single system characterised by
a spin temperature $\approxgt$10$^3$~K (\cite{ostorero09}).
We report a possible anti-correlation
between the projected linear size of the radio source
and the X-ray column density, analogous to the anti-correlation
between radio size and HI column density reported by
Pihlstr\"om et al. (2003) and Gupta et al. (2006)

\item GPS galaxies occupy a specific locus in the
radio versus X-ray luminosity plane. They lie well on the
extrapolation to high radio powers of
the correlation between these two quantities discovered in
low-luminosity FR~I radio galaxies. On the
other hand, GPS galaxies exhibit a comparable X-ray
luminosity to FR~II radio galaxies, notwithstanding their
much greater radio luminosity

\item GPS galaxies occupy the same locus as Seyfert
galaxies in the O[III] to X-ray luminosity ratio versus
X-ray column density diagnostic plane

\end{itemize}

If the bulk of the X-ray emission in GPS galaxies is
due to the accretion disk, one may expect that
the X-ray and the radio source evolution in GPS galaxies
would be decoupled. The accretion disk would
maintain a stable flow of gas and (mostly sublimated)
dust to the supermassive
black hole, while the radio source fades away with its
expansion into the ISM. This would imply an {\it increase}
of the X-ray to radio luminosity ratio with size. On the
other hand, models of the dynamical evolution of GPS sources,
where the X-ray emission is primarily produced by Compton
upscattering of ambient photons (\cite{stawarz08}), predict
a {\it decrease} of the X-ray to radio luminosity ratio with
size. The $L_X/L_{5 GHz}$ ratios observed in our sample are
consistent with both possibilities. Measurements in
$\gamma$-rays by {\it Fermi} (as originally proposed by
\cite{stawarz08}) could be crucial to discriminate
between them. Alternatively, the large scatter observed
in this quantity may be indicative of a coupling between
the accretion disk and the jet activity driven by
disk instabilities. 

The evolutionary scenarios described above postulate
that GPS sources are young objects, as also indicated by
the direct measurement of their dynamical age.
Eventually GPS galaxies would
reach their full maturity as classical FR~II radio galaxies.
However, column densities $\approxgt$10$^{22}$~cm$^{-2}$
fully surrounding the expanding radio source could significantly
brake, if not entirely inhibit, this state, leading
to a significant underestimate of dynamical ages based on
hotspot recession velocity measurements.
Extending the number of X-ray measurements of
low-luminosity ($L_{5 \ GHz} =
10^{40-42}$~erg~s$^{-1}$) GPS galaxies is the next step we
intend to pursue, in order to pinpoint the endpoint of their
evolution.

\section*{Appendix~A: a serendipitous blazar in the field of 4C+00.02}

A bright off-axis source is visible in the EPIC field of view of the
XMM-\textit{Newton} 4C+00.02 observation. This source will be referred to in
the following as
XMMU~J002200.8+000655. It is outside the field of view of the
XMM-Newton Optical Monitor.
The best-fit parameters when the baseline model is
applied to its spectrum are summarised in Table~\ref{modeldat}. 
% ---------------- Tab.1 - Appendix
\begin{table}
\caption{\label{modeldat}Fitting results for XMMU~J002200.8+000655}
\begin{center}
\begin{tabular}{cccc}
\hline 
$\mathrm{N_H}$$^a$ & $\mathrm{\Gamma}$ & $\mathrm{N}$ & $\mathrm{\chi^{2}_{\nu}}/\nu$ \\
($\mathrm{10^{20}}$~cm$^{-2}$) &     &  (10$^{-4}$~cm$^{-2}$~s$^{-1}$)   \\ \hline
$4.7^{+1.5}_{-1.6}$ & $2.29^{+0.05}_{-0.06}$ & $7.40 \pm 0.04$ & 1.00/350 \\ \hline
\end{tabular}
\end{center}

\noindent
$^a$in excess of the Galactic contribution along the line-of-sight to
XMMU~J002200.8+000655.
\end{table}
% ---------------- Tab.1 - Appendix
The EW of a unresolved Fe K$_{\alpha}$ neutral fluorescent line is
constrained to be lower than 400~eV.

\begin{figure}
\centering
\includegraphics[width=9.5cm,angle=90]{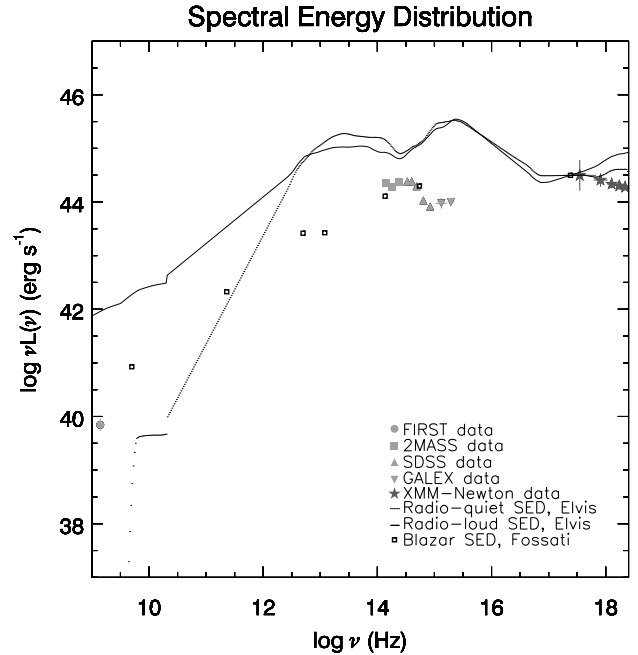}
\caption{The Spectral Energy Distribution (SED) for XMMU~J002200.8+000655.
The photometric points
are compared with standard SED for Blazars (boxes; \cite{fossati98}) and
radio-quiet and
radio-loud quasars (dotted line and solid line, respectively; \cite{elvis94a}),
normalised at the value of XMMU~J002200.8+000655 at 1~keV.}
\label{SED}
\end{figure}

This source was already known as 1RXS~J002200.9+000659
(\cite{voges99}). We have searched with
{\sc Aladin} for available measurements at other wavelengths.
We found data in GALEX GR4,
SDSS (\cite{adelman08}), 2MASS,
and FIRST (\cite{white97}) results. In Fig.~\ref{SED} we compare the
Spectral Energy Distribution (SED), with the average SED for radio-quiet and
radio-loud quasars normalised at 1~keV
(\cite{elvis94b}) and with the ``blazar track''
corresponding to objects with the same radio luminosity as
XMMU~J002200.8+000655
(\cite{fossati98}).
As can be seen in Fig.\ref{SED}
the blazar SED track qualitatively agrees with the XMMU~J002200.8+000655 SED.
%----------------- Fig.1 Appendix 
%----------------- Fig.1 Appendix 
This source appears in NED as a BL Lac candidate, and in
V\'eron-Cetty \& Veron (2006) as a confirmed BL Lac object.
Its optical spectrum was previously studied by Collinge et al.
(2005),
but heavy contamination by the host galaxy prevented
its properties from being properly studied. 

\section*{Appendix~B: summary of the radio and X-ray properties of the whole GPS sample}

In Table~\ref{tabap1} we report a summary of the X-ray properties of the
whole GPS sample discussed in this paper, together with the 5~GHz
luminosity after O'Dea (1998) and Stanghellini et al. (1998).
% ---------------- Tab.1 - Appendix B
\begin{table*}
\caption{X-ray properties and radio luminosity of the whole GPS sample
discussed in this paper (Sect.~5 and 6)}
\label{tabap1}
\begin{tabular}{lcccccc} \hline \hline
Source name & $\log (L_{\rm 5 \ GHz}, {\rm erg \ s^{-1}})$ & $L_{{\rm 2-10 \ keV}}$ & $N_H$ & Size & HI $\log({\rm N_H, cm^{-2}}$) & $\Gamma$ \\
 & & ($10^{43}$~erg~s$^{-1}$) & ($10^{20}$~cm$^{-2}$) & (kpc) & & \\ \hline
4C+00.02       & 43.1 & $0.55$$^{\dag}$           & $<10000$                & 0.220  & ...      & ... \\ 
COINSJ0111+3906& 44.0 & $7 \pm 3$        & $>500$                  & 0.033  & 21.9     & ... \\
PKS0428+20     & 43.1 & $1.4 \pm 0.6$    & $<70$                   & 0.653  & 20.5     & ... \\
PKS0500+019    & 43.6 & $50 \pm 6$       & $50 \pm^{18}_{16}$      & 0.055  & 20.8     & $1.61\pm^{0.16}_{0.15}$ \\
B30710+439     & 43.8 & $34.0 \pm 0.2$   & $51 \pm^{17}_{14}$     & 0.118  & ...      & $1.59\pm0.06$ \\
PKS0941$-$080    & 42.8 & $0.091 \pm 0.075$& $<10000$                & 0.148  & $<$20.1  & ... \\
B1031+567      & 43.4 & $2.2 \pm 0.2$    & $50 \pm 18$             & 0.206  & $<$20.1  & ... \\
4C+14.41       & 43.2 & $1.40 \pm 0.19$  & $<16$                   & 0.306  & $<$19.8  & ... \\
4C+32.44       & 43.6 & $3.7 \pm 0.4$    & $12 \pm^6_5$            & 0.247  & 19.9     & $1.7 \pm 0.2$ \\ 
PKS1345+125    & 42.7 & $7.8 \pm 1.5$    & $480 \pm 40$            & 0.166  & 20.6     & $1.1\pm^{0.7}_{0.8}$ \\
4C+62.22       & 43.6 & $30 \pm 20$      & $290 \pm^{200}_{100}$   & 0.218  & 20.3     & $1.24\pm0.17$ \\
OQ+208         & 42.2 & $>9.0$           & $>9000$                 & 0.010  & 20.3     & $2.21\pm^{0.19}_{0.14}$ \\
PKS1607+26     & 43.8 & $29.0 \pm 0.1$   & $>6000$                 & 0.262  & ...      & ... \\
PKS2008-068    & 44.5 & $ 6 \pm 2$       & $<50$                   & 0.218  & ...      & ... \\
PKS2127+04     & 44.6 & $29 \pm 4$       & $<19$                   & 0.218  & ...      &  $2.0\pm^{0.5}_{0.4}$ \\
COINSJ2355+4950& 43.0 & $1.3 \pm 0.3$    & $400 \pm^{700}_{300}$   & 0.199  & 20.5     & $1.8\pm^{1.6}_{0.9}$ \\
\hline \hline
\end{tabular}

\noindent
$^{\dag}$extrapolated from the detection in the 0.5--2~keV band with $\Gamma = 1.61$ and $N_H = N_{\rm H,Gal}$
\end{table*}
% ---------------- Tab.1 - Appendix B

\begin{acknowledgements}

Based on observations obtained with XMM-Newton, an ESA science mission
with instruments and contributions directly funded by ESA Member States
and NASA This research has made use of
data obtained through the High Energy Astrophysics Science Archive
Research Centre Online Service, provided by the NASA/Goddard Space
Flight Centre and of the NASA/IPAC Extragalactic Database (NED) which
is operated by the Jet Propulsion Laboratory, California Institute of
Technology, under contract with the National Aeronautics and Space
Administration.
OT thanks the whole ESA administration, and in particular
Nienke de Boer, Marcus Kirsch and Fernando Maura, for their
support during a six-month traineeship at ESAC, where most
of the data analysis included in this paper was performed.
This research is funded in part by NASA grant NNX07AQ55G.
OT gratefully acknowledge an ESA Internal Fellowship Trainee grant. AS
is partly supported by NASA contract NAS8-39073.
The authors thank C.Stanghellini
for a critical revision of an early version of this manuscript.
Last, but not least, the authors gratefully acknowledge a careful and accurate
referee report by Dr.D.~J.~Saikia, which greatly improved the
overall presentation of the paper, while allowing us
to better clarify some aspects of the radio measurements discussed in
this paper.

\end{acknowledgements}

\end{document}